\documentclass[aps,twocolumn,showpacs,superscriptaddress,nofootinbib]{revtex4}  % for review and submission
\usepackage{graphics,graphicx}  % needed for figures
\usepackage{dcolumn}   % needed for some tables
\usepackage{bm}        % for math
\usepackage{amssymb}   % for math
\usepackage{bigints}
\usepackage{xcolor}
\usepackage[toc,page]{appendix}
\def\be{\begin{equation}}
\def\ee{\end{equation}}
\def\bea{\begin{eqnarray}}
\def\eea{\end{eqnarray}}

\begin{document}

\widetext

\title{Hydrogen Bond of QCD in Doubly Heavy Baryons and Tetraquarks
}

\author{Luciano Maiani}
\affiliation{T. D. Lee Institute, Shanghai Jiao Tong   University, Shanghai, 200240, China}
\affiliation{Dipartimento di Fisica and INFN,  Sapienza  Universit\`a di Roma, Piazzale Aldo Moro 2, I-00185 Roma, Italy}
\author{Antonio D. Polosa}
\affiliation{Dipartimento di Fisica and INFN,  Sapienza  Universit\`a di Roma, Piazzale Aldo Moro 2, I-00185 Roma, Italy}
\author{Veronica Riquer}
\affiliation{T. D. Lee Institute, Shanghai Jiao Tong   University, Shanghai, 200240, China}
\affiliation{Dipartimento di Fisica and INFN,  Sapienza  Universit\`a di Roma, Piazzale Aldo Moro 2, I-00185 Roma, Italy}
\email{luciano.maiani@roma1.infn.it}
\email{antonio.polosa@roma1.infn.it}
\email{veronica.riquer@cern.ch}

\date{\today}

\begin{abstract}
 In this paper we present in greater detail previous work on the Born-Oppenheimer approximation to treat the hydrogen bond of QCD, and add a similar treatment of doubly heavy baryons. Doubly heavy exotic resonances $X$ and $Z$ can be described as color molecules of two-quark lumps, the analogue of the $H_2$ molecule, and doubly heavy baryons as the analog of the $H_2^+$ ion, except that the two heavy quarks attract each other. We compare our results with constituent quark model and lattice QCD calculations and find further evidence in support of this upgraded picture of compact tetraquarks and baryons.    
\end{abstract}

\pacs{12.40.Yx, 12.39.-x, 14.40.Lb}
\maketitle

\section{ Introduction}
Systems with heavy and light particles allow for an approximate treatment  where the light and heavy degrees of freedom are studied separately, and solved one after the other. This is the Born-Oppenheimer approximation (BO), introduced in non-relativistic Quantum Mechanics for molecules and crystals, where electrons coexist with the much heavier nuclei. We have recently reconsidered  this method for the QCD interactions of multiquark hadrons containing heavy (charm or bottom) and light (up and down)  quarks~\cite{noiprd}, following earlier work in ~\cite{braatenBO,Brambilla:2017uyf}, and, for lattice calculations, in~\cite{bicudo}.

In this paper, based on our previous communication~\cite{noiprd}, we  consider tetraquarks in terms of {\it color molecules}: lumps of two-quark colored atoms ({\it orbitals})  held together  by color forces and treated in the Born-Oppenheimer (BO) approximation. 
The variety of bound states described here identifies a new way of looking at multiquark hadrons, as formed by the QCD analog of the hydrogen bond of molecular physics. 

We restrict to doubly heavy-light systems, namely the doubly heavy baryons, $qQQ$, not considered in~\cite{noiprd},   the hidden flavor tetraquarks $Q\bar Q q \bar q$, see~\cite{Maiani:2004vq,Maiani:2014aja,book,Esposito:2016noz}  for a review,  and $QQ\bar q\bar q$ systems~\cite{Esposito:2013fma,Karliner:2017qjm,Eichten:2017ffp,Eichten:2017ual,Luo:2017eub}.
 
The plan of the paper is the following. 

Sect.~\ref{BOapp} describes the Born-Oppenheimer approximation applied to a QCD double heavy hadron and gives the two-body color couplings derived from the restriction that the hadron is an overall color singlet. 
Sect.~\ref{m&s} recalls the salient features of the constituent quark model and gives quark masses and hyperfine couplings derived from the mass spectra of the $S$-wave mesons and baryons.
Sect.~\ref{q_int} introduces the string tension for confined systems and discusses extensions beyond charmonium. 

Sects.~\ref{doubch},  \ref{cbarctetra}, \ref{bbtetra} illustrate the main calculations and results for doubly heavy baryons, 
 hidden heavy flavor and doubly heavy flavored tetraquarks, respectively. 

Results are summarised in Sect.~\ref{summ} and conclusions given in Sect.~\ref{concl}. Technical details are expanded in three Appendices.

\section{Born-Oppenheimer approximation with QCD constituent quarks }\label{BOapp}

We consider doubly heavy systems with open or hidden heavy flavor, and discuss  the application of the Born-Oppenheimer (BO) approximation along the lines used for the treatment of the hydrogen molecule, see~\cite{weinbergQM,pauling}. 

We denote coordinates and mass of the heavy quarks by ${\bm x}_A, {\bm x}_B$ and $M$ and those of the light quarks by  ${\bm x}_1, {\bm x}_2$ and $m$. Coordinate symbols  include here spin and color quantum numbers, to be discussed later.

The hamiltonian of the whole system is
\bea
H&=&\frac{1}{2M}\sum_{\rm heavy} P_i^2 +\frac{1}{2m} \sum_{\rm light} p_i^2 +\nonumber \\
&+&V({\bm x}_A, {\bm x}_B)+V_{I}({\bm x}_A, {\bm x}_B, {\bm x}_1, {\bm x}_2)\label{compl}
\eea
We have separated the heavy quark interaction $V({\bm x}_A, {\bm x}_B)$, {\it e.g.} their Coulombic QCD interaction, from the general interactions involving light-heavy and light-light quarks. 

We start by solving the eigenvalue equation for the light particles for fixed values of the coordinates of the heavy ones
\be
\left( \sum_{\rm light} \frac{p_i^2}{2m} +V_{ I}({\bm x}_A, {\bm x}_B, {\bm x}_1, {\bm x}_2)\right)f_\alpha={\cal E}_\alpha({\bm x}_A, {\bm x}_B)f_\alpha 
\label{redbo}
\ee
where
\be
f_\alpha=f_\alpha({\bm x}_A, {\bm x}_B, {\bm x}_1, {\bm x}_2) 
\ee
and focus on the lowest eigenvalue and eigenfunction, which, dropping the subscript for simplicity of notation,  we denote by ${\cal E}$ and $f$. Next, we look for solutions of the eigenvalue equation of the complete Hamiltonian~\eqref{compl} of the form
\be
\Psi=\psi({\bm x}_A, {\bm x}_B) f({\bm x}_A, {\bm x}_B, {\bm x}_1, {\bm x}_2)
\ee
When applying the Hamiltonian~\eqref{compl} to $\Psi$ one encounters terms 
of the kind
\bea
&&P_{A,B} \Psi=\psi({\bm x}_A, {\bm x}_B)~i\frac{\partial}{\partial {\bm x}_{A,B}}f({\bm x}_A, {\bm x}_B, {\bm x}_1, {\bm x}_2)+\nonumber \\&&\qquad +\left[i\frac{\partial}{\partial{\bm x}_{A,B}} \psi({\bm x}_A, {\bm x}_B)\right]f({\bm x}_A, {\bm x}_B, {\bm x}_1, {\bm x}_2)
~\label{deriv}
\eea
The Born-Oppenheimer approximation consists in neglecting the first with respect to the second term in all such instances 
so that, after factorizing $f$, we obtain the Schr\"odinger equation  of the heavy particles
\be
\left(\sum_{\rm heavy} \frac{P_i^2}{2M}+V_{BO}({\bm x}_A, {\bm x}_B)\right)\psi=E\psi  \label{bofin}
\ee
with the Born-Oppenheimer potential given by
\be
V_{\rm BO}({\bm x}_A, {\bm x}_B)= V({\bm x}_A, {\bm x}_B) + {\cal E}({\bm x}_A, {\bm x}_B) \label{bopot}
\ee

For QED in molecular physics, the parameter which regulates the validity of the approximation is estimated in~\cite{weinbergQM}  to be
\be
\epsilon= \left(\frac{m}{M}\right)^{1/4}
\ee
We apply the same method to our case as follows. 

The ratio of the first (neglected) to the second (retained) term in \eqref{deriv} is given approximately by
\be
\Lambda=\frac{1/a}{1/b}
\ee 
where $a$ and $b$ are the lengths over which $f$ or $\psi$ show an appreciable variation. 

The length $a$ is simply the radius of the orbitals, which we determine by minimizing the Schr\"odinger functional of the light quark. As will be discussed below, we find typically $1/a=A\sim0.3$~GeV, {\it i.e.} $a\sim0.7$~fm.

The length $b$ has to be formed from the  dimensional quantities over which the Born-Oppenheimer equation \eqref{bofin} depends. In the case of double heavy baryons and hidden heavy flavor tetraquarks, Sects.~\ref{doubch} and~\ref{cbarctetra}, Eq.~\eqref{bofin} depends on $1/M$, on $a$ and on the string tension $k$, which has dimensions of GeV$^2$.

A quantity $b$ with dimensions of length can be formed as
\be
b=\left(M k A\right)^{-1/4}
\ee

Therefore 
\be
\Lambda=A^{3/4}(k M)^{-1/4}\label{err1}
\ee
which is 0.57 for charm and 0.43 for beauty,
using $k=0.15$~GeV$^2$ and the constituent masses of charm and beauty from the Tables in the next Section.
 
 We note in Sect~\ref{bbtetra} that the Born-Oppenheimer potential for double heavy tetraquarks does not depend on the string tension, which is screened by gluons for color octet orbitals. In this case, we get
 \be
b=\left(M  A\right)^{-1/2}
\ee
 and 
  \bea
  \Lambda=\left(\frac{A}{M}\right)^{1/2}\label{err2}
\eea
giving 0.42 for charm and 0.24 for beauty.
In the following, for convenience we shall include quark masses in $V_{BO}$, but it is worth noticing that the error we are estimating is the error on the binding energies, which turn out to be  around $100$~MeV or smaller  in absolute value. So, the errors corresponding to~\eqref{err1} and \eqref{err2} may be in the order of $20-50$~ MeV.

We comment now about color. Treating heavy quark and/or antiquark as external sources implies specifying their combined $SU(3)_c$ representation. Restriction to an overall color singlet fixes completely the color composition of the constituents. 

Recall that the color coupling between any pair of particles in color representation ${\bm R}$ is given by
 \bea 
 &&V_{C}(r)=\lambda_{q_1 q_2}({\bm R})~\frac{\alpha_s}{r}\nonumber \\
 &&\lambda_{q_1 q_2}({\bm R})=\frac{1}{2}[C_2({\bm R})-C_2(\bm q_1)-C_2(\bm q_2)]\label{casimir1}
 \eea
where $\bm q_{1,2}$ are the irreducible representations of the particles in the pair and $C_2$ the quadratic Casimir operators. 

We note the results:
$C_2({\bm 1})=0$; $C_2({\bm R})=C_2({\bar{\bm R}})$; $C_2 ({\bm 3})=4/3$; $C_2({\bm 6})=10/3$;  $C_2({\bm 8})=3$.

If the pair $q_1q_2$ in the tetraquark $T(q_iq_jq_kq_l)$ is in a superposition of two $SU(3)_c$ representations with amplitudes $a$ and $b$, we use
\bea
&&T= a\,|(q_1 q_2)_{{\bm  R}_1}\dots \rangle_{\bm 1}+b\,|(q_1 q_2)_{{\bm  R}_2}\dots \rangle_{\bm 1}\nonumber  \\
&&\lambda_{q_1q_2 }=a^2\lambda_{q_1 q_2}({\bm R_1})+b^2\lambda_{q_1 q_2}({\bm R_2})\label{casimir2}
\eea

The different cases are as follows.

{\bf \emph{Doubly charmed baryon: $\bm c\bm c$ in ${\bar {\bm 3}}$}}.
In a color singlet baryon, all pairs are in color ${\bar{\bm 3}}$, and the color couplings are distributed according to
\be
\lambda_{c c}=\lambda_{c q}= -2/3 \label{lambar}
\ee

{\bf \emph{Hidden flavor tetraquarks.}}
Color of the heavy particles can be either ${\bm 1}$ or ${\bm 8}$. 
In the first case, the interaction between $Q\bar Q$ and $q\bar q$ pairs goes via the exchange of color singlets.  We are in a situation  dominated by nuclear-like forces, eventually leading to the formation of the {\it hadrocharmonium} envisaged in~\cite{Dubynskiy:2008mq}. We shall not consider $Q\bar Q$ in color singlet any  further.

{\bf \emph{${\bm Q}{\bar{ \bm Q}}$ in ${\bm 8}$}}.
Suppressing coordinates
\be
T=(\bar Q \lambda^A Q)(\bar q \lambda^A q) \label{tetra}
\ee
with the sum over $A=1,\dots,8$ understood. If we restrict to one-gluon exchange, Eq.~\eqref{tetra} determines the interactions between different pairs.

Both $Q\bar Q$ and $q\bar q$ are in color octet and we read their coupling from Eq.~\eqref{casimir1}. The couplings of the other pairs are found using the Fierz rearrangement formulae for SU(3)$_c$ to bring the desired pair in the same quark bilinear (see Appendix~\ref{fierz}). We get in total
 \bea
 &&\lambda_{c\bar c}=\lambda_{q\bar q}= +\frac{1}{6} \nonumber\\
 &&\lambda_{cq}=\lambda_{\bar c\bar q}= -\frac{1}{3}\label{lamccbar}\\
&&\lambda_{c\bar q}=\lambda_{\bar c q}= -\frac{7}{6}\nonumber
 \eea

 Substituting  light and heavy quarks with electrons and protons, respectively, we see that the pattern of repulsions and attractions given by Eqs.~(\ref{lamccbar}) is the same as that of the hydrogen molecule.
   
{\bf \emph{Double beauty tetraquarks: $\bm b\bm b$ in ${\bar{\bm 3}}$}}.
The lowest  energy state corresponds to $ bb$ in spin one and light antiquarks in spin and isospin zero. 
The tetraquark state 
\be T=|(bb)_{\bar {\bm 3}}, (\bar q\bar q)_{ {\bm 3}} \rangle_{\bm 1}\label{tetra3}
\ee
can be Fierz transformed into
\be
T=\sqrt{\frac{1}{3}}|(\bar q b)_{\bm 1},(\bar q b)_{\bm 1}\rangle _{\bm 1}-\sqrt{\frac{2}{3}}|(\bar q b)_{\bm 8},(\bar q b)_{\bm 8}\rangle _{\bm 1}
\ee
with all attractive couplings
\be
\lambda_{bb}=\lambda_{\bar q \bar q}=-\frac{2}{3}\alpha_S\quad 
\lambda_{b\bar q}
=-\frac{1}{3}\alpha_S\label{bqbar3}
\ee

{\bf \emph{Double beauty tetraquarks: $\bm b\bm b$ in $\bm 6$}}.
We start from 
\be 
T=|(bb)_{\bm 6}, (\bar q\bar q)_{{\bar {\bm 6}}}\rangle_{\bm 1} \label{tetra6}
\ee
a case also considered in~\cite{Luo:2017eub}. We find
\be
T=\sqrt{\frac{2}{3}}|(\bar q b)_{\bm 1},(\bar q b)_{\bm 1}\rangle _{\bm 1}+\sqrt{\frac{1}{3}}|(\bar q b)_{\bm 8},(\bar q b)_{\bm 8}\rangle _{\bm 1}
\ee
therefore
\be
\lambda_{bb}=\lambda_{\bar q \bar q}=+\frac{1}{3}\alpha_S
\quad \lambda_{b\bar q}
=-\frac{5}{6}\alpha_S\label{bqbar6}
\ee
The situation is again analogous to the $H_2$ molecule, with two identical, repelling light particles. 
%@@@@@@@@@@@@@@@@@@@@@@@@@@@@@@@@@@@@@@@@@

\section{Quark masses and hyperfine couplings from Mesons and Baryons}\label{m&s}

The constituent quark model, in its simplest incarnation, describes the masses of mesons and baryons as due to the masses of the quarks  in the hadron, $M_i$, with hyperfine interactions  added. The Hamiltonian is
\bea
&&H=H_{\rm mass}+H_{\rm hf}\nonumber \\
&&H_{\rm mass}=\sum_i M_i\\
&&H_{\rm hf}=\sum_{i<j} 2\kappa_{ij}({\bm s}_i\cdot{\bm s}_j)\nonumber 
\eea
where ${\bm s}$ is the constituent's spin and $H_{\rm hf}$ denotes the hyperfine interaction term.

This picture gives a reasonable description of the masses of uncharmed, single charm and single beauty mesons, with four well determined quark masses. It gives an equally reasonable description of baryon masses, albeit with a set of slightly different quark masses, as shown in  Tab.~\ref{mas}.

\begin{table}[htb!]
\centering
    \begin{tabular}{|c|c|c|c|c|}
     \hline
Quark Flavors& q &  s &  c & b \\ \hline
Quark mass (MeV) from mesons & $308$ & $484$ & $1667$ & $5005$ \\ \hline
Quark mass (MeV) from baryons & $362$ & $540$ & $1710$ & $5044$ \\ \hline
\end{tabular}
 \caption{\footnotesize {Constituent quark masses (MeV) from $S$-wave mesons and baryons, see~\cite{book,Maiani:2004vq}, ($q=u,d$).}}
\label{mas}
\end{table}

\begin{table}[htb!]
\centering
    \begin{tabular}{|c|c|c|c|c|c|c|c|} 
    \hline
Mesons & $(q\bar q)_1$&$(q\bar s)_1$&  $(q\bar c)_1$&  $(s\bar c)_1$ & $(q \bar b)_1$& $(c\bar c)_1$ & $(b\bar b)_1$  \\ 
\hline
$\kappa$ (MeV) & $318$ & $200$  & $70$ & $72$ & $23$  & 56 &   30\\ 
\hline\hline
Baryons & $(qq)_{\bar 3}$ & $(q s)_{\bar 3}$  &  $(q c)_{\bar 3}$ &  $(s c)_{\bar 3}$& $(q b)_{\bar 3}$& $(c c)_3$ & $(b b)_3$   \\ 
\hline
$\kappa$ (MeV) & $98$ &$59$ &   $15$ & $50$ & $2.5$ & 28\footnote{~$0.5\kappa[(c\bar c)_1]$} & 15\footnote{~$0.5\kappa[(b\bar b)_1]$}   \\ 
\hline
 \hline
Ratio $\frac{\kappa_{MES}}{\kappa_{BAR}}$ & 3.2& 3.4  & 4.7 &1.6  & 9.2 & -- &-- \\
\hline
\end{tabular}
 \caption{\footnotesize {$S$-wave Mesons and Baryons: spin-spin interactions of the lightest quarks with the heavier flavors~\cite{book,Maiani:2004vq}. For  the hf couplings of $c\bar c$, $c c$ and similar ones for $b$ quarks see Text.}}
 \label{spin}
\end{table}

Values for $\kappa[(c\bar c)_1]$ are taken from the mass differences of ortho- and para-quarkonia, 
{\it e.g.} $\kappa[(c\bar c)_1]=1/2(M_{J/\psi}-M_{\eta_c})$. Those for $\kappa[(cc)_3]$ and $\kappa[(b b)_3]$ are obtained multiplying by the one-gluon exchange color factor 1/2.

A reasonable hypothesis, advanced in~\cite{Karliner:2014gca}, is that the difference of  quark masses derived from mesons and baryons is due to the different pattern of QCD interactions in systems with two or three constituents, that should be apparent even in the lowest order, one-gluon exchange approximation. 
We shall follow this hypothesis. 
Since the basic ingredient of the BO approximation are two body orbitals, we feel the natural choice is {\it to take quark masses from the meson spectrum} and leave to the QCD interactions between orbitals the task to compensate for the difference of quark masses from mesons with those derived from baryons in the naive constituent quark model.

\section{Quark interaction and string tension}\label{q_int}

The prototype of non relativistic quark interaction is the so-called Cornell potential~\cite{cornell} introduced in connection with charmonium spectrum,
 \be
V(r)=-\frac{4}{3}~ \frac{\alpha_S}{r} + k r+V_0=V_{C}(r)+V_{\rm conf}(r)+V_0~\label{quarkpot}
\ee
The potential refers to the case of a heavy color triplet pair, $Q\bar Q$, in an overall color singlet state. $V_0$ is determined from the mass spectrum. We shall generalise~\eqref{quarkpot} to several different cases.

The first term in~\eqref{quarkpot}  is obtained in the one-gluon exchange  approximation by \eqref{casimir1}. It is generalised to any pair of colored particles in a color representation ${\bm R}$ by Eq.~\eqref {casimir2}.

The second term in \eqref{quarkpot}, which dominates over the Coulomb force at large separations, arises from quark confinement. 
In the simplest picture, confinement is due to the condensation of Coulomb lines of force into a string that joins the quark and the antiquark. The linearly rising term in \eqref{quarkpot} describes the force transmitted by the string tension. In this picture, it is natural to assume that the string tension, embodied by the coefficient $k$,
scales with the Coulomb coefficient
\be
k_{q_1 q_2}\propto |\lambda_{q_1 q_2}|   \label{scaling}
\ee
For color charges combined in an overall color singlet, the assumption leads to $k\propto C_2({\bf q})$, whence the name of Casimir scaling, see~\cite{Bali:2000gf} for an extensive discussion. 

Casimir scaling would give a string tension that increases with the dimensionality of color charges. However, QCD gluons, unlike photons in QED, may screen color charges, by lowering the dimension of the color representation. The simplest case is color ${\bm 6}$ charge. Since ${\bm 6}\otimes{\bm 8}\supset  {\bar{\bm 3}}$ the string tension strength of a pair ${\bm 6}\otimes{\bar{\bm 6}}\to {\bm 1}$ is reduced from $10/3$ to $4/3$,  {\it i.e.} the string tension of ${\bm 3}\otimes{\bar{\bm 3}}\to{\bm 1}$~\cite{Bali:2000gf}.

Screening by an arbitrary number of gluons reduces Casimir scaling of string tension to the simplest {\it triality scaling}. One can see this through the following steps.
\begin{enumerate}
\item Recall that a generic $SU(3)_c$ charge is represented by a tensor $t_{a\dots}^{b\dots}$ ($a,~b,\dots=1,2,3$ are $SU(3)_c$ indices), with $n$ upper and $m$ lower, fully symmetrized indices  and vanishing under contraction of an upper and a lower index~\cite{coleman}.  Exchanging $n$ and $m$ gives the complex conjugate representation, which has the same Casimir, so that we may assume $n\geq m$;  $t=(n-m)~{\rm mod}~3=0,1,2$ is the {\it triality} of the representation.
\item Saturating the tensor $t$ with gluon fields: $t_{a\dots}^{b\dots}~A^a_b\dots$ we reduce to tensors which have only $n-m$  upper indices;
\item ${\bm 8}\otimes{\bm 8}^\prime\supset {\bm 1}{\bm 0},{\overline{{\bm 1}{\bm 0}}}$ as can be seen from the simple composition of two (different) octets
\be
G_{\{abc\}}=[(A)_a^d(A^\prime)_b^e\epsilon_{cde}]_{{\rm abc~symmetrized}}\in {\overline{\bm{10}}}
\ee

\item saturating $t$ with $G_{\{abc\}}$ we reduce the upper indices to those of the lowest triality representations, namely $({\bm 1}, t=0)$;  $({\bm 3}, t=1)$; $({\bm 6}, t=2)$, equivalent to~\footnote{Indeed $t=(n-m)~{\rm mod}~3 \equiv (n-m)-3\lfloor\frac{n-m}{3}\rfloor$} $({\bar{\bm 3}}, t=-1)$.
\end{enumerate} 
The upshot is that, for conjugate charges combined to a singlet,  we have only two possibilities for the string tension: 
\begin{itemize}
\item$t=0$ charges, {\it e.g.} ${\bm 8}\otimes{\bm 8}\to {\bm 1}$, have $k=0$ and are not confined 
\item  $t\neq 0$ charges have $k=k({\bm 3})$, equal to the charmonium string tension. 
\end{itemize}
For other cases, {\it e.g.} $qQ$, we adopt \eqref{scaling} and write
\be
V_{q_1 q_2}(r)=\lambda_{q_1 q_2}~\frac{\alpha_S}{r}+\frac{3|\lambda_{q_1 q_2}|}{4}~k r + V_0 \label{ptgen}
\ee
where $k$ is the string tension taken from charmonium spectrum.

In the numerical applications, 
we take $\alpha_S$ and $k$ at the charmonium scale from the lattice calculation in~\cite{lattice}:
\be
\alpha_S(2M_c)=0.30\quad\quad k=0.15~{\rm GeV}^2 \label{cparam}
\ee
At the $B_c$ meson and bottomonium mass scales we take the same string tension and run $\alpha_S$ with the two loops beta function, to get:
\be
\alpha_S(M_c+M_b)=0.24\quad\quad \alpha_S(2M_b)=0.21 \label{bb&bc}
\ee

\section{The Doubly charmed baryon}\label{doubch}

The baryon $\Xi_{cc}=qcc$ is analogous to the $H_2^+$ ion~\cite{pauling}, except that the two heavy quarks attract each  other, Eq.~\eqref{lambar}.

The interaction Hamiltonian is
\bea
&&H_I=-\frac{2}{3} \alpha_S~\frac{1}{|{\bm x}_A-{\bm x}_B|}+\nonumber \\
&&-\frac{2}{3}\alpha_S~\left(\frac{1}{|{\bm x}-{\bm x}_A|}+\frac{1}{|{\bm x}-{\bm x}_B|}\right)\label{ccbarhtot}
\eea

We consider the orbital made by $cq$, with $c$ located in ${\bm x}_A$. The perturbation Hamiltonian that remains is
\be
H_{\rm pert}=-\frac{2}{3}\alpha_S ~\frac{1}{|{\bm x}-{\bm x}_B|}
\label{bcc}
\ee

{\bf \emph{The ${\bm c}{\bm q}$ orbital.}}
As potential, we take the Coulombic interaction from \eqref{ccbarhtot} and  a linear term with  the string tension rescaled according to Eq.~\eqref{ptgen}, 
\be
V_{cq}=-\frac{2}{3}\frac{ \alpha_S}{r}+\frac{1}{2}k\, r + V_0 \label{dhorbit}
\ee

We assume a radial wave-function $R(r)$ of the form
\be
R(r) =\frac{A^{3/2}}{\sqrt{4\pi}}~e ^{-Ar}\label{radwf}
\ee
and determine $A$ by minimizing the Schr\"odinger  functional
\be
\langle H(A)\rangle=\frac{\Big(R(r),(-\frac{1}{2M_q}\Delta+V_{cq}-V_0)R(r)\Big)}{(R(r),R(r))}\label{schroedfun}
\ee
We take quark masses from the meson spectrum, Tab.~\ref{mas}, $\alpha_S$ and $k$ from~\eqref{cparam}. We find
$
A=0.32~{\rm GeV}, \, \langle H\rangle_{\rm min} = 0.48~{\rm GeV} 
$.

We consider as unperturbed ground state the symmetric superposition of the two orbitals with $q$ attached either to $c({\bm x}_A)$, which we denote by $\psi({\bm x})$, or attached to $c({\bm x}_B)$,  denoted by $\phi({\bm x})$. 
\be 
f ({\bm x})=\frac{\psi({\bm x})+\phi({\bm x})}{\sqrt{2(1+S)}}=\frac{R(|{\bm x}-{\bm x}_A|)+R(|{\bm x}-{\bm x}_B|)}{\sqrt{2(1+S)}}\label{groundbar}
\ee
The denominator in~\eqref{groundbar} is needed to normalise $f({\bm x})$ and it arises because $\psi$ and $\phi$ are not orthogonal, see Appendix~\ref{qedmol}, with the overlap $S$ defined as ($\psi$ and $\phi$ real):
\be
S(r_{AB})=\int d^3\xi~\psi(\xi)\phi(\xi)\label{overlap}
\ee
 and $r_{AB}=|{\bm x}_A-{\bm x}_B|$.
 
The energy corresponding to $f({\bm x})$ is given by the quark constituent masses plus the energy of the orbital
\be
E_0= 2M_c+M_q + \langle H\rangle_{\rm min} +V_0\label{unpertenB}
\ee

{\bf \emph{Perturbation theory.}}
To first order in the perturbation~\eqref{bcc}, the BO potential is given by
\be
V_{\rm BO}(r_{AB})=-\frac{2}{3}\alpha_S\frac{1}{r_{AB}}+ E_0+\Delta E(r_{AB}) 
\ee

\bea 
&&\Delta E(r_{AB})=\langle f| H_{\rm pert}|f\rangle=\notag\\
&&=-\frac{2\alpha_S}{3}\frac{1}{1+S}\left[I_1(r_{AB})+I_2(r_{AB})\right]
\label{deltae0}
\eea
$I_{1,2}$ are functions of $r_{AB}$ defined in terms of $\psi$ and $\phi$:
\bea
&&I_1(r_{AB})= \int d^3\xi\,|\psi(\xi)|^2 \, \frac{1}{|{\bm \xi}-{\bm x}_B|}\label{i1} \\
&&I_2(r_{AB})= \int d^3\xi\,\psi(\xi)\phi(\xi)~ \frac{1}{|{\bm \xi}-{\bm x}_B|} \label{i2}
\eea
where the vector $\bm \xi$ originates from $A$, taken in the origin, and $|\bm x_B|=r_{AB}$.
 
 Analytic expressions for $S,I_{1,2}$ are given in \cite{pauling} for the hydrogen wave functions. We evaluate  them numerically for the orbitals corresponding to the potential~\eqref{dhorbit}.

{\bf \emph{Boundary condition for ${\bm r_{AB}\to 0}$}}.
The perturbation Hamiltonian~\eqref{bcc} embodies the interaction of the  light quark when the other  charm quark is far from the orbital. If we  let $r_{AB}$ to vanish, the charm pair reduces to a single ${\bar{\bm 3}}$ source generating the same interaction that $q$ would see inside a $q{\bar c}$ charmed meson. This is in essence the heavy quark-diquark symmetry, see~\cite{Savage:1990di, Brambilla:2005yk,Fleming:2005pd}. 

The symmetry means that $E_0+\Delta E(r_{AB})$, when we subtract $M_c$ from it and let $r_{AB}\to 0$, has to reproduce the spin independent mass of a $\bar D$ meson, which, by definition, is $M_c+M_q$. In formulae
\bea
&&E_0+\Delta E(0) -M_c=\nonumber \\
&&=M_c+M_q +V_0+\langle H\rangle_{\rm min} +\Delta E(0)=\nonumber \\
&&=M_c+M_q
\eea
The condition determines the value of the a priori unknown $V_0$
\be
V_0+\langle H\rangle_{\rm min}+\Delta E(0)=0\label{boundcond}
\ee
and
\bea
&&V_{\rm BO}(r_{AB})= -\frac{2}{3}\alpha_S\frac{1}{r_{AB}}+ \Delta E(r_{AB})+C\nonumber \\
&&C=2M_c+M_q  -  \Delta E(0)
\label{VBOcorr}
\eea

Numerically, we find from~\eqref{deltae0}
\be
\Delta E(0)=-65~{\rm MeV}
\ee

{\bf \emph{The ${\bm {cq}}$ orbital is confined}}.
The interactions embodied in Eq.~\eqref{VBOcorr} originate from one gluon exchange and vanish at large separations. However, the orbital $cq$ and the external $c$ quark carry ${\bar {\bm 3}}$ and  $ {\bm 3}$ colors combined to a color singlet and are confined.  To take this into account we add a linearly rising term to the BO potential in \eqref{VBOcorr}, determined by the string tension $k$ of charmonium, see Sect.~\ref{q_int}, and the onset point, $R_0$. The complete  Born-Oppenheimer potential reads
\bea
&&V_{\rm tot}(r)=V_{\rm BO}(r) +V_{\rm conf}(r)\label{BOtot} \\
&& V_{\rm conf}(r)=k \times (r-R_0)\times \theta(r-R_0)\label{conf}
\eea
with $R_0\geq 2 A^{-1}$. For orientation, we start with $R_0\sim 8$~GeV$^{-1}\sim 1.6~$fm, where we may assume that $c$ sees the orbital as a  point source and study the results for different values of  $R_0$.

\begin{figure}[htb!]
\begin{minipage}[c]{4.5cm}
 %  \centering
 %\begin{center}
   \includegraphics[width=4.3truecm]{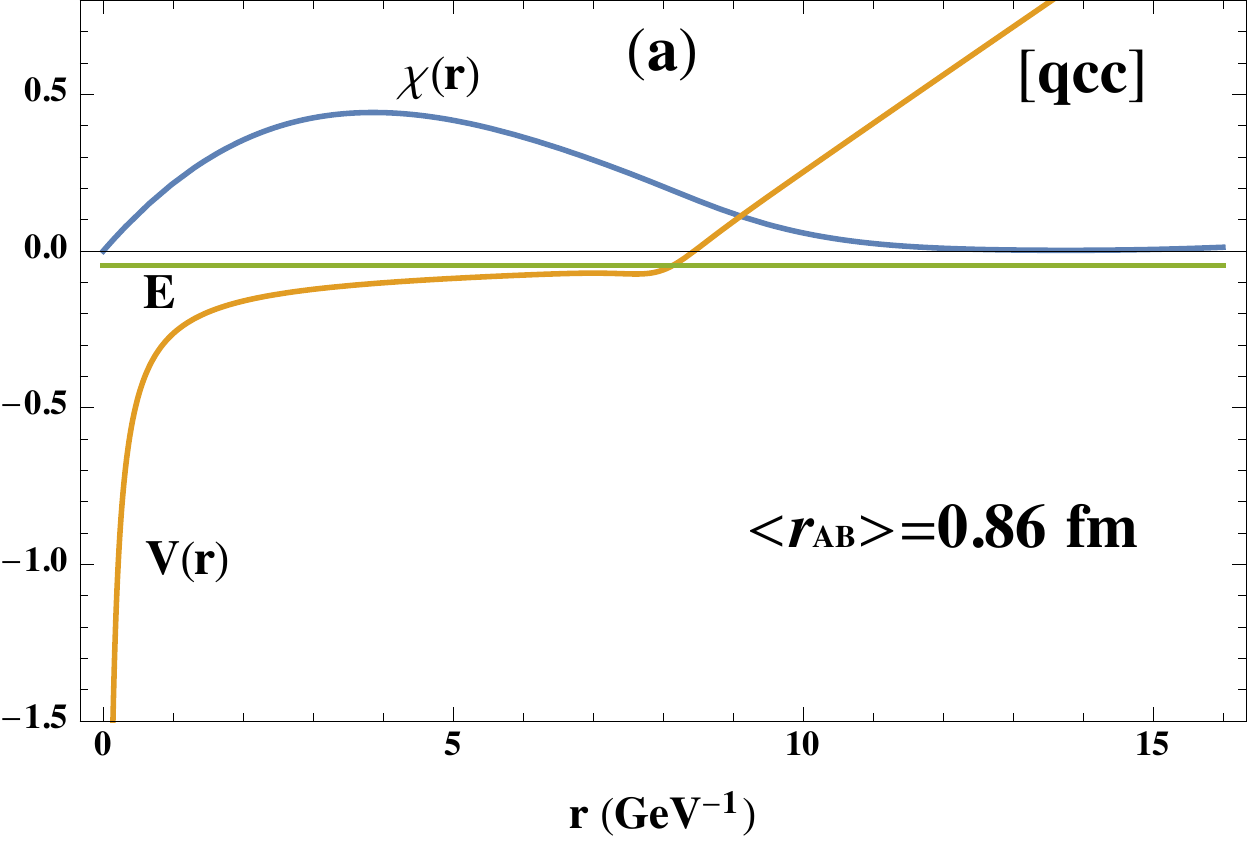}
   \end{minipage}%
% \end{center}
 \begin{minipage}[c]{4.5cm}
%\centering
  \includegraphics[width=4.3truecm]{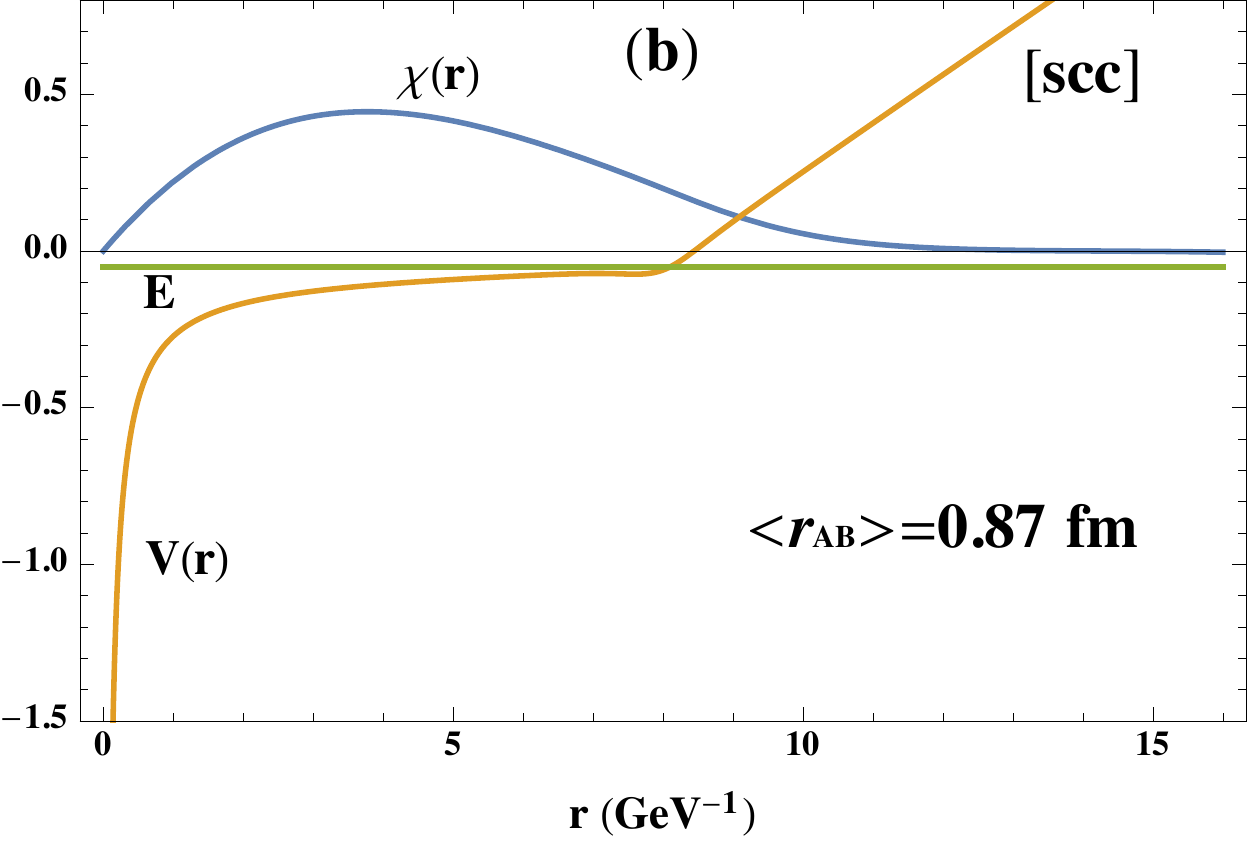}
\end{minipage}%
\caption{\footnotesize Born-Oppenheimer potential + confinement  in the $qcc$ (a) and $scc$ (b) baryons.
Eigenfunction $\chi(r)=rR(r)$ and eigenvalue $E$  in the fundamental state are shown. Here and in the following, on  the y-axes energies are in GeV and $\chi$ in arbitrary units.
\label{ccbarav}}
\end{figure}

The Schr\"odinger equation for the charm pair with potential $V(r)=V_{\rm tot}(r)-C$ is solved numerically~\cite{schroed}. Results are reported in Fig.~\ref{ccbarav}. For $R_0=8$~GeV$^{-1}$, we plot $V(r)$, the radial wave function $\chi(r)$ and the lowest eigenvalue $E=-0.041$~GeV. The average distance of the charm pair is $\sim 0.9$~fm. The eigenvalue  has an appreciable dependence from $R_0$. We find
\be
E= -41^{+17}_{-7}~{\rm MeV}~{\rm for}~R_0=8\pm 2 ~{\rm GeV}^{-1}
\ee
The contribution of  hyperfine interactions to the $J=1/2^+$ $\Xi_{cc}$ is
\be
H_{\rm hf}(\Xi_{cc})=-2\kappa_{qc}+\frac{1}{2}\kappa_{cc}=-16~{\rm MeV}
\ee
with the numerical value from Tab.\ref{spin}. Finally
\bea
&&M(\Xi_{cc})_{{\rm Th}} = \nonumber\\
&&=2M_c+M_q -\Delta E(0) + E -2\kappa_{qc}+\frac{1}{2}\kappa_{cc}\label{mxicc}
\eea
leading to
\be
M(\Xi_{cc})_{{\rm Th}}=3655^{+17}_{-7}\label{xicc}~{\rm MeV}
\ee
to be compared with the LHCb value~\cite{Aaij:2018gfl}
\be
M(\Xi_{cc})_{{\rm Expt}}=3621.2\pm 0.7~{\rm MeV}
\ee

We do not attempt to give an overall theoretical error to the result in~\eqref{xicc}, which cannot be however less than $\pm 30$ MeV.

It is interesting to compare our with the calculation presented in~\cite{Karliner:2014gca}. These authors obtain the $c\bar c$ binding energy from charmonium using quark masses from the meson spectrum (particle names denote their masses in MeV)
\be
B_{c\bar c}=\frac{1}{4}(\eta_c+3 J/\psi)-2M_c=-271
\ee
where the first term is charmonium mass subtracted of its hyperfine interaction. The $cc$ binding energy is obtained by multiplication of the color factor $1/2$, and the result is used as binding energy of the $cc$ quarks in $\Xi_{cc}$, to be subtracted from $cc$ quark mass derived from the baryon spectrum. Adding hyperfine interactions, they obtain~\cite{Karliner:2014gca}:
\bea
&&\Xi_{cc} %[{\rm K\&R}]
=3628\pm 12
\eea

The consistency of results derived by two alternative routes with themselves and with the experimental value is worth noticing.

%@@@@@@@@@@@@@@@@@@@@@@@@@@@@@@@@@@@@@@@
\begin{table}[htb!]
\centering
    \begin{tabular}{|c|c|c|c|c|c|}
     \hline
-- &   $- \Delta E(0)$ & $E$ &  $M[\Xi_{QQ}] $& \cite{Karliner:2014gca,Karliner:2018hos} & \cite{Mathur:2018rwu,Mathur:2018epb}\\ \hline 
$\Xi_{cc}$   & $+65$   &  $-41$  &  $3656$ & $3628\pm 12$ & $3634(20)$\\ \hline 
$\Omega_{cc}$   & $+75$   &  $-44$  &  $3769$ & $3692 \pm16$ & $3712(11)(12)$\\ \hline \hline
$\Xi_{cb}$  & $+50$   &  $-37$  &  $6961$& $6920\pm 13$ & $6945(22)(14)$ \\ \hline
$\Xi_{cb}^\prime$   & $+50$   &  $-37$   &$6993$ &$6935\pm 12$  &     $6966(23)(14)$        \\ \hline \hline
$\Xi_{bb}$    & $+44$   &  $-53$  &  $10311$& $10162\pm 12$ & --- \\ \hline 
\end{tabular}
 \caption{\footnotesize Our results on doubly-heavy baryon masses, fourth column, compared to quark model and lattice QCD results, fifth and sixth columns. $E-\Delta E$ represents the correction to the constituent quark mass formula, 
with quark masses taken from meson spectrum.
}
\label{dhbar}
\end{table}
 \emph{{$\bm{ \Omega{cc}}$.}}
Replacing the light  quark mass with the strange quark mass in \eqref{schroedfun} and inserting the appropriate hyperfine couplings, we  obtain the mass of the strange-doubly charmed baryon, $[scc]$, denoted by $\Omega_{cc}$.

{\bf \emph{Mass of ${\bm {cb}}$ and ${\bm {bb}}$ baryons.}}
With similar methods we may compute $M[\Xi_{cb}]$, $M[\Xi_{cb}]^\prime$, see Appendix~\ref{xispin}, and $M[\Xi_{bb}]$. 

{\bf \emph{Comparisons.}}
Our results are summarized in Tab.~\ref{dhbar}, fourth column and compared to the results in Ref.~\cite{Karliner:2014gca,Karliner:2018hos}, reported in the fifth column. We differ for $bc$ and $bb$ by 50 and 150 MeV, which perhaps points to a significant discrepancy.

Predictions of the masses of doubly heavy baryons, based on different methods,  have appeared earlier in the literature~\cite{Bjorken:1986xpa,Anikeev:2001rk,Richard:1994ae,Roncaglia:1995az,Ebert:1996ec,Kiselev:2001fw,Narodetskii:2002ib,He:2004px,Albertus:2006ya,Roberts:2007ni,Gerasyuta:2008zy,Weng:2010rb,Zhang:2008rt}. Numerical values are summarized in~\cite{Karliner:2014gca} and spread in a typical range of 100-200 MeV around our values.

The results of recent lattice QCD calculations~\cite{Mathur:2018epb,Mathur:2018rwu} are reported in the last column. Ref.~\cite{Mathur:2018epb} reviews the results of today available lattice calculations for doubly heavy baryons.

Experimental results are eagerly awaited.

\section{Hidden Heavy Flavor}\label{cbarctetra}

We consider the hidden heavy flavor case, specializing to hidden charm and following closely the approach to the $H_2$ molecule in~\cite{pauling}, see Appendix~\ref{qedmol}. 

With $c\bar c$ and $q\bar q$ taken in color $\bm 8$ representation,  Eq.~\eqref{tetra}, we describe the unperturbed state as the product of two orbitals,  bound states of one heavy and one light particle around ${\bm x}_A$ or ${\bm x}_B$, and treat the interactions not included in the orbitals as perturbations.

Two subcases are allowed: $i)$  $cq$ and $\bar c \bar q$ or $ii)$ $c\bar q$ and $\bar c q$.  

{\bf \emph{The ${\bm c}{\bm q}$ orbital.}}
We take the Coulombic interaction given by $\lambda_{cq}$ in \eqref{lamccbar} and rescale the string tension from the charmonium one, according to Eq.~\eqref{ptgen}, thus\footnote{In our previous analysis~\cite{noiprd}, string tension $1/4 k$ was considered as an alternative possibility to string tension $k$.}
\be
\label{orbitpot}
V_{cq}=-\frac{1}{3}\frac{ \alpha_S}{r}+\frac{1}{4}kr + V_0
\ee

Like the previous case, we assume an exponential form for radial wave-function $R(r)$
\be
R(r) =\frac{A^{3/2}}{\sqrt{4\pi}}~e ^{-Ar}\label{radwf}
\ee
and determine $A$ by minimizing the Schroedinger  functional~\eqref{schroedfun} for the potential~\eqref{orbitpot}, with quark masses from the meson spectrum, 
Tab.~\ref{mas}, and parameters of the potential from~\eqref{cparam}. 
We find
$
A=0.27~{\rm GeV}, \, \langle H\rangle_{\rm min} = 0.30 ~{\rm GeV} 
$.

The wave function of the two non interacting orbitals is 
\be 
f(1,2)=\psi(1)\phi(2)=R(|{\bm x}_1-{\bm x}_A|)R(|{\bm x}_2-{\bm x}_B|)
\ee
Unlike the $H_2$ case, light particles are not identical and the unperturbed  ground state is non degenerate.

The energy of $f(1,2)$ is given by 
\be
E_0= 2(M_c+M_q + \langle H \rangle_{\rm min} +V_0)\label{unperten}
\ee

{\bf \emph{Perturbation theory.}}
The perturbation Hamiltonian of this case is:
\bea
H_{\rm pert}&=&-\frac{7}{6}\alpha_S \left(\frac{1}{|{\bm x}_1-{\bm x}_B|}+\frac{1}{|{\bm x}_2-{\bm x}_A|}\right) +\nonumber \\
&&+\frac{1}{6}\alpha_S\,\frac{1}{|{\bm x}_1-{\bm x}_2|}\label{pertccbar}
\eea

 To first order in $H_{\rm pert}$, Eq.~\eqref{pertccbar}, the BO potential is 
\be
V_{\rm BO}(r_{AB})=+\frac{1}{6}\alpha_S\frac{1}{r_{AB}}+ E_0+\Delta E(r_{AB}) 
\ee
where $r_{AB}=|{\bm x}_A-{\bm x}_B|$. 

 $\Delta E= \langle  f  |H_{\rm pert}| f \rangle$ evaluates to
\be
\Delta E = -\frac{7}{6}\alpha_S \, 2 I_1(r_{AB})+
\frac{1}{6}\alpha_S\,  I_4(r_{AB})\label{vbopert}
\ee
in terms of the  function $ I_1$, Eq.~\eqref{i1}, and 
\be
I_4(r_{AB})= \int d^3\xi d^3\eta\, |\psi(\xi)|^2\,|\phi(\eta)|^2 \frac{1}{|{\bm \xi}-{\bm \eta}|} \label{i4}
\ee
where the vector $\bm \xi$ originates from $A$, taken in the origin, and $|\bm x_B|=r_{AB}$.

{\bf \emph{${\bm {cq}}$ orbitals are confined}.}
The orbitals $cq$ and $\bar c\bar q$ carry non vanishing color and are confined.  Similarly to Sect.~\ref{doubch}, we add a linearly rising term to the BO potential in \eqref{vbopert}, determined by a string tension $k_T$ and the onset point, $R_0$. The complete  Born-Oppenheimer potential reads
\bea
&&V(r)=V_{\rm BO}(r) +V_{\rm conf}(r)\label{BOtot}\nonumber \\
&& V_{\rm conf}(r)=k_T \times (r-R_0)\times \theta(r-R_0)
\eea

For orientation, we choose $R_0=10$~GeV$^{-1}$, greater than $2A^{-1}\sim 7.4$ GeV$^{-1}$, where the two orbitals start to separate. In principle, $R_0$ should be considered a  free parameter, to be fixed on the phenomenology of the tetraquark,  as we discuss below.

 As for $k_T$, we note that
the tetraquark $T=|(\bar c c)_{\bm 8} (\bar q q)_{\bm 8}\rangle_{\bm 1}$ can be written as
\be
T=\sqrt{\frac{2}{3}} |(cq)_{\bar{\bm 3}} (\bar c \bar q)_{\bm 3}\rangle_{\bm 1}-\sqrt{\frac{1}{3}}|(cq)_{\bm 6} (\bar c \bar q)_{{\bar {\bm 6}}}\rangle_{\bm 1}\label{3&6}
\ee
At large distances the diquark-antidiquark system is a superposition of ${\bar {\bm 3}}\otimes {\bm 3}\to {\bm 1}$ and ${\bm 6}\otimes{\bar {\bm 6}} \to {\bm 1}$. 
Eq.~\eqref{3&6} and the hypothesis of strict Casimir scaling of $k_T$ would give
\be
k_T=\left(\frac{2}{3}+\frac{1}{3}~ \frac{C_2({\bm 6})}{C_2({\bm 3})}\right) k=1.5~k
\ee
However, as discussed in~\cite{Bali:2000gf} and in Sect.~\ref{q_int}, gluon screening gives the ${\bm 6}$ diquark a component over the ${\bar{\bm 3}}$ bringing $k_T$ closer to $k$. For simplicity, we adopt $k_T=k$.

\begin{figure}[htb!]
\begin{minipage}[c]{4cm}
   \centering
 %\begin{center}
   \includegraphics[width=4truecm]{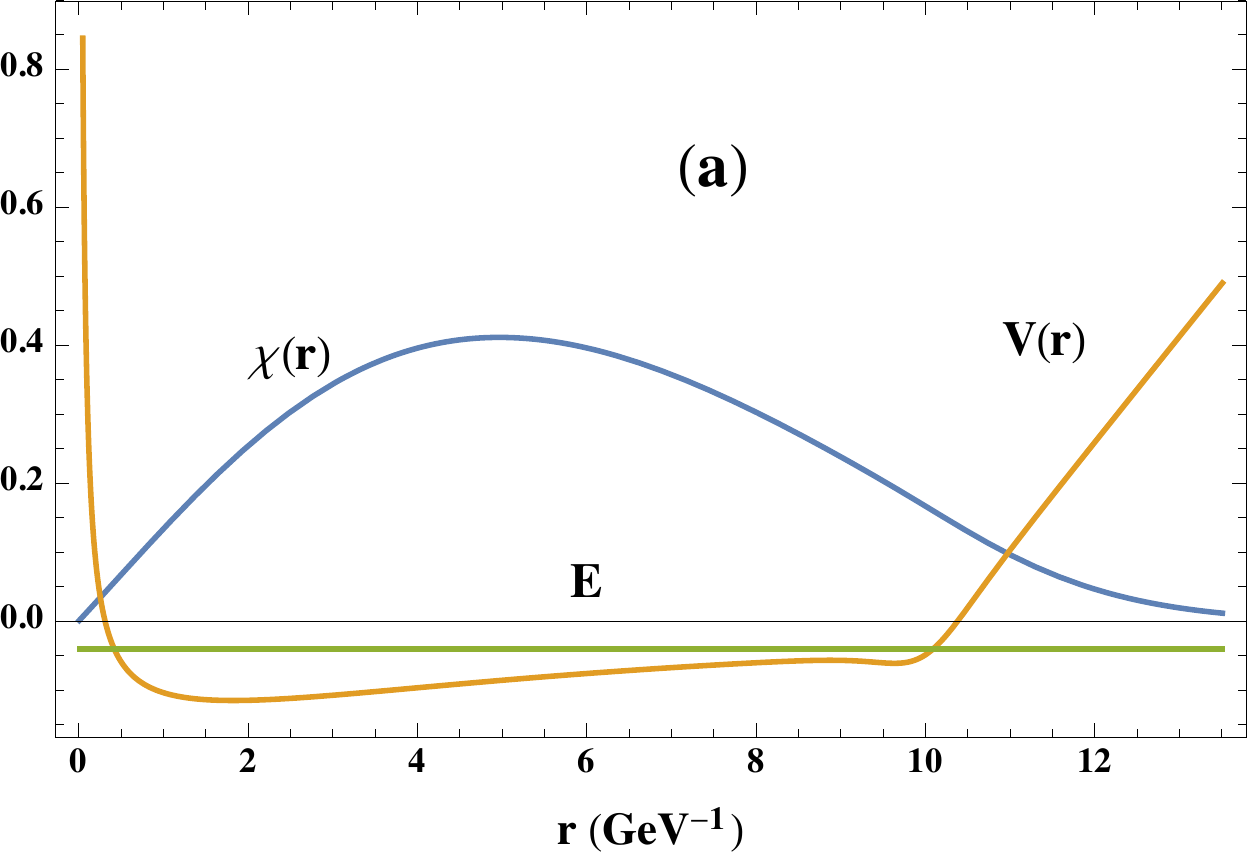}
    \end{minipage}%
% \end{center}
 \begin{minipage}[c]{5cm}
\centering
   \includegraphics[width=4truecm]{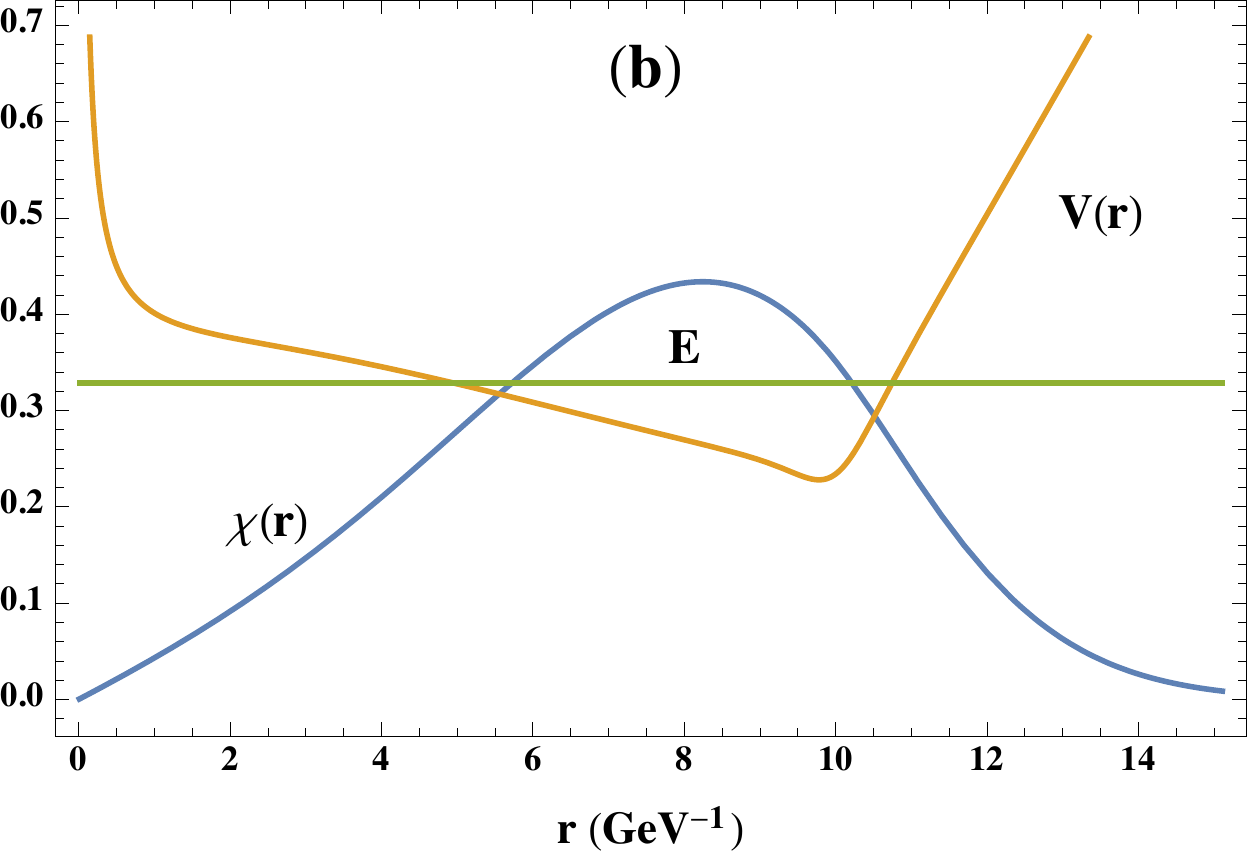}
 \end{minipage}%
\caption{\footnotesize (a) dominant $c\bar q$ and $\bar c q$ attraction + confinement; (b) dominant $q \bar q$ repulsion + confinement, letting  $+1/6 \,\alpha_S\sim 0.05 \to 3.3$ in Eq.~\eqref{lamccbar}. Eigenfunction $\chi(r)=rR(r)$ and eigenvalue $E$ of the tetraquark in the fundamental state are shown. Diquarks are separated by a potential barrier and there are two different lenghts: $R_{qc}\sim 0.4 $~fm and the total radius $R\sim 1.5$~fm, as in~\cite{Maiani:2017kyi}. 
\label{uno}}
\end{figure}

The  potential $V(r)$ computed on the basis of Eqs.~(\ref{BOtot}) is given in Fig.~\ref{uno}(a). 
Also reported are the wave function and the eigenvalue obtained by solving numerically the radial Schr\"odinger 
equation~\cite{schroed}.

As it is customary for confined system like charmonia, we fix $V_0$ to reproduce the mass of the tetraquark, so the eigenvalue is not interesting. However, the eigenfunction gives us information on the internal configuration of the tetraquark. 
In Fig.~\ref{uno}(a), with one-gluon exchange couplings, a configuration with $c$ close to $\bar c$ and the light quarks around is obtained, much like the quarkonium adjoint meson described in~\cite{braatenBO}. 

Fig.~\ref{uno}(b) is obtained by increasing the repulsion in the $q\bar q$ interaction associated to the function $I_4$, letting $1/6\,\alpha_S= 0.05 \to 3.3$. 
The corresponding $c\bar c$ wave function clearly displays the separation of the diquark from the antidiquark suggested in~\cite{Maiani:2017kyi} and further considered in~\cite{Esposito:2018cwh}. 

The presence of a barrier that $c$ has to overcome to reach $\bar c$, apparent in Fig.~\ref{uno}(b), explains the suppression of the $J/\psi+ \rho/\omega$ decay modes of $X(3872)$, otherwise favored by phase space with respect to the $D D^*$ modes. 
With the parameters in Fig.~\ref{uno}(b), we find $|R(0)|^2=1.6\cdot10^{-3}$ 
~with respect to $|R(0)|^2=1.9\cdot 10^{-2}$ with the perturbative parameters of Fig.~\ref{uno}(a).

%@@@@@@@@@@@@@@@@@@@@@@@@@@

\begin{figure}[htb!]
\begin{minipage}[c]{4cm}
   \centering
 %\begin{center}
   \includegraphics[width=4.05truecm]{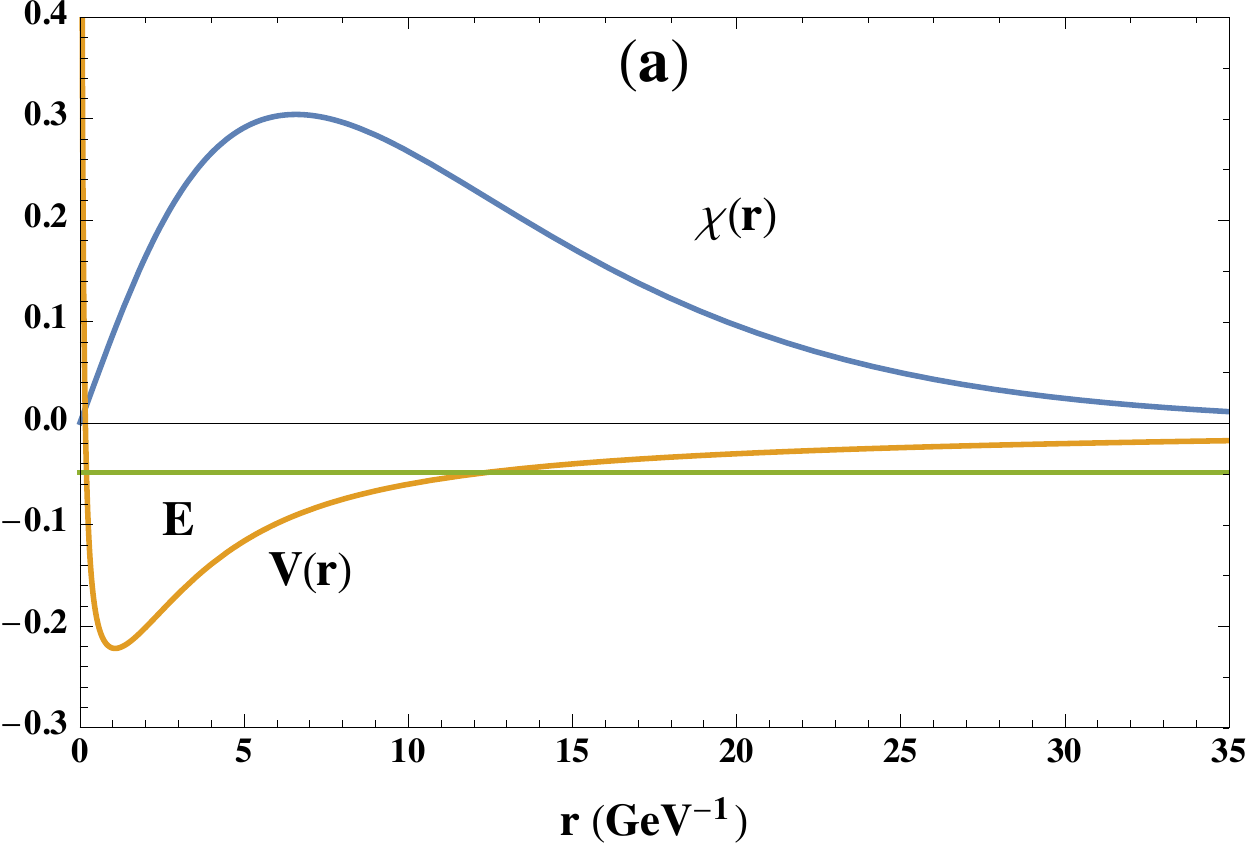}
    \end{minipage}%
% \end{center}
 \begin{minipage}[c]{5cm}
\centering
   \includegraphics[width=4truecm]{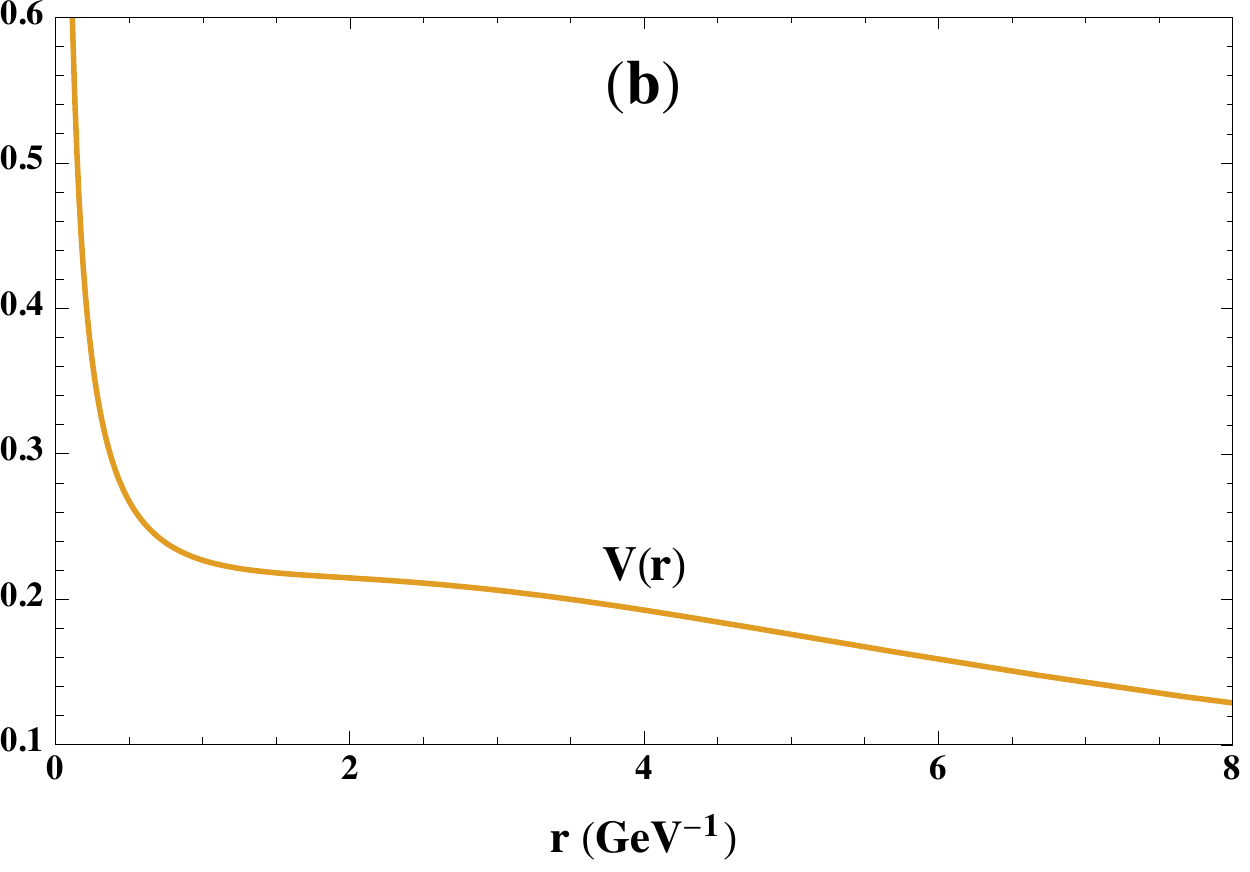}
 \end{minipage}%
\caption{\footnotesize {Born-Oppenheimer potential $V(r)$ vs. $R_{AB}$ for $c\bar q$ orbitals. Unit lenght: GeV$^{-1}\sim 0.2$~fm.   (a) using the perturbative parameters;%, $E=-17$~MeV;
~(b) with repulsion enhanced.}
}
\label{due}
\end{figure}

The tetraquark picture of $X(3872)$ and the related $Z(3900)$ and $Z(4020)$ have been originally formulated  in terms of pure ${\bar {\bm 3}}\otimes {\bm 3}$ diquark-antidiquark states~\cite{Maiani:2004vq,Maiani:2014aja,Maiani:2017kyi}.
The ${\bm 6}\otimes{\bar {\bm 6}}$ component in (\ref{3&6}) results in the opposite sign of the $q\bar q$ hyperfine interactions vs the dominant $c q$ and $\bar c \bar q$ one, and it could be the reason why $X(3872)$ is lighter than $Z(3900)$.

{\bf \emph{The ${\bm c}{\bar{\bm q}}$ orbital.}}
One obtains the new orbital by replacing  $-1/3~\alpha_S \to -7/6~\alpha_S$ in Eq.~(\ref{orbitpot}) and string tension
\be
k(c\bar q)=\frac{7}{8}\, k
\ee
 Correspondingly
$
A=0.40~{\rm GeV},\, \langle H\rangle_{\rm min}  = 0.66~{\rm GeV}
$.
The perturbation Hamiltonian appropriate to this case is
\bea
H_{\rm pert}
&=&-\frac{1}{3}~\alpha_S \left(\frac{1}{|{\bm x}_1-{\bm x}_B|}+\frac{1}{|{\bm x}_2-{\bm x}_A|}\right) +\nonumber \\
&&+\frac{1}{6}~\alpha_S~\frac{1}{|{\bm x}_1-{\bm x}_2|} \label{pertcqbar}
\eea
and
\be
V_{\rm BO}=+\frac{1}{6}~\alpha_S ~\frac{1}{r_{AB}}+ E_0 +\Delta E(r_{AB}) \label{qbarcpot}
\ee
with 
\be
\Delta E= -\frac{1}{3}\alpha_S \, 2 I_1+
\frac{1}{6}\alpha_S\,  I_4  \label{decqbar}
\ee

The tetraquark state is 
\be
T=\sqrt{\frac{8}{9}}|(\bar c q)_{\bm 1}(\bar q c)_{\bm 1}\rangle_{\bm 1}-\frac{1}{\sqrt{9}}|(\bar c q)_{\bm 8}(\bar q c)_{\bm 8}\rangle_{\bm 1} \label{cqbar}
\ee

At large $|{\bm x}_A-{\bm x}_B|$ the lowest energy state
(two color singlet mesons) has to prevail, as concluded in Sect.~\ref{q_int} on the basis of the triality scaling due to gluon screening of octet charges. Therefore there is no confining potential to be added to the BO potential in \eqref{qbarcpot}.

{\bf \emph{Boundary condition for ${\bm r_{AB}\to \infty}$.}}
 For $r_{AB}\to \infty$, $V_{\rm BO}\to \langle H\rangle_{\rm min} +V_0$. Including constituent quark masses, the energy of the state at $r_{AB}= \infty$ is  
$
E_\infty=2(M_c + M_q+ \langle H\rangle_{\rm min} +V_0)
$ 
and it must coincide with the mass of %the two orbitals corresponding to 
a pair of non-interacting charmed mesons, with spin-spin interaction subtracted. 
Therefore we impose
\be
\langle H\rangle_{\rm min} +V_0=0
\label{noconf}
\ee
A minimum of the BO potential is not guaranteed. If there is such a minimum, as in Fig.~\ref{due}(a), it would correspond to a configuration similar to the quarkonium adjoint meson in Fig.~\ref{uno}(a). 

If repulsion is increased above the perturbative value,  {\it e.g.} changing $+1/6 ~\alpha_S\sim 0.05$ to a coupling $\geq 1$ in analogy with Fig.~\ref{uno}(b), the BO potential has no minimum at all, Fig.~\ref{due}(b).

\section{Double beauty tetraquarks.} \label{bbtetra}
We consider $bb$ tetraquarks, analyzing in turn the two options for the total color of the $b b$ pair.

$\bm b\bm b$ in ${\bar{\bm 3}}$.
We recall from Sect.~\ref{BOapp} that the lowest  energy state corresponds to $ bb$ in spin one and light antiquarks in spin and isospin zero. The tetraquark state is $T=|(bb)_{\bar {\bm 3}}, (\bar q\bar q)_{ {\bm 3}} \rangle_{\bm 1}$, whence one derives the attractive color couplings reported in \eqref{bqbar3} and
\be
k(b\bar q)=\frac{1}{4} k \label{kbqbar3}
\ee
There is only one possible orbital, namely $b\bar q$, but the unperturbed state now is the superposition of two states with the roles of $\bar q_1$ and  $\bar q_2$ interchanged, like electrons in the $H_2$ molecule, see Appendix~\ref{qedmol}.
\be
f(1,2)=\frac{\psi(1)\phi(2)+\phi(1)\psi(2)}{\sqrt{2\left(1+S^2 \right)}}
\ee
The denominator needed to normalise $f(1,2)$ includes the overlap function $S$ defined in \eqref{overlap}. 

The perturbation Hamiltonian is
\bea
H_{\rm pert}&=&-\frac{1}{3}~\alpha_S \left(\frac{1}{|{\bm x}_1-{\bm x}_B|}+\frac{1}{|{\bm x}_2-{\bm x}_A|}\right) +\nonumber \\
&&-\frac{2}{3}~\alpha_S~\frac{1}{|{\bm x}_1-{\bm x}_2|}
\eea
and
\be
V_{\rm BO}(r_{AB})=2 (\langle H\rangle_{\rm min} +V_0)-\frac{2}{3}\alpha_S\frac{1}{r_{AB}} + \Delta E
\ee
where $\Delta E=\langle f|  H_{\rm pert}| f\rangle$ evaluates to
\be
\Delta E=\frac{1}{1+S^2}\left[ -\frac{1}{3}\alpha_S ~2( I_1+S I_2)-\frac{2}{3}\alpha_S(I_4 + I_6)\right]\label{bb3}
\ee
$I_{1,2,4}$ were defined previously whereas~\cite{pauling}
\bea
%&& I_6(r_{AB})=\nonumber \\
 I_6(r_{AB})=\int d^3\xi d^3\eta\, \psi(\xi)\phi(\xi) \psi(\eta)\phi(\eta)\frac{1}{|{\bm \xi}-{\bm \eta}|}
\label{i6}
\eea
For the orbital $b\bar q$ we find
$
A=0.26~{\rm GeV},\, \langle H\rangle_{\rm min} = 0.32~{\rm GeV}
%\label{orbitbb3}    
$.
%@@@@@@@@@@@@@@@@@@@@@@@@@@@@@@@@@@@@@@@@@
\begin{figure}[htb!]
\begin{minipage}[c]{4.2cm}
   \centering
 %\begin{center}
   \includegraphics[width=4.2truecm]{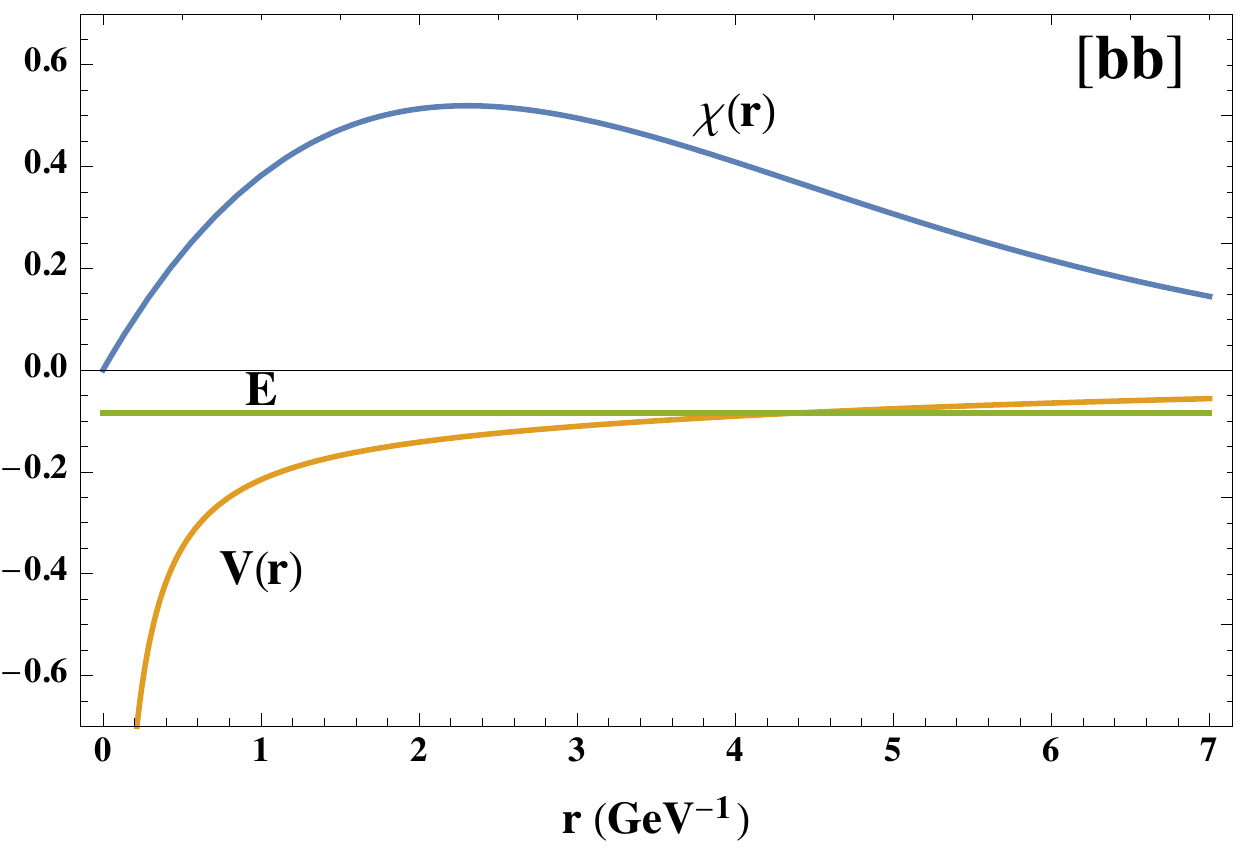}
    \end{minipage}%
% \end{center}
 \begin{minipage}[c]{4.8cm}
\centering
   \includegraphics[width=4.26truecm]{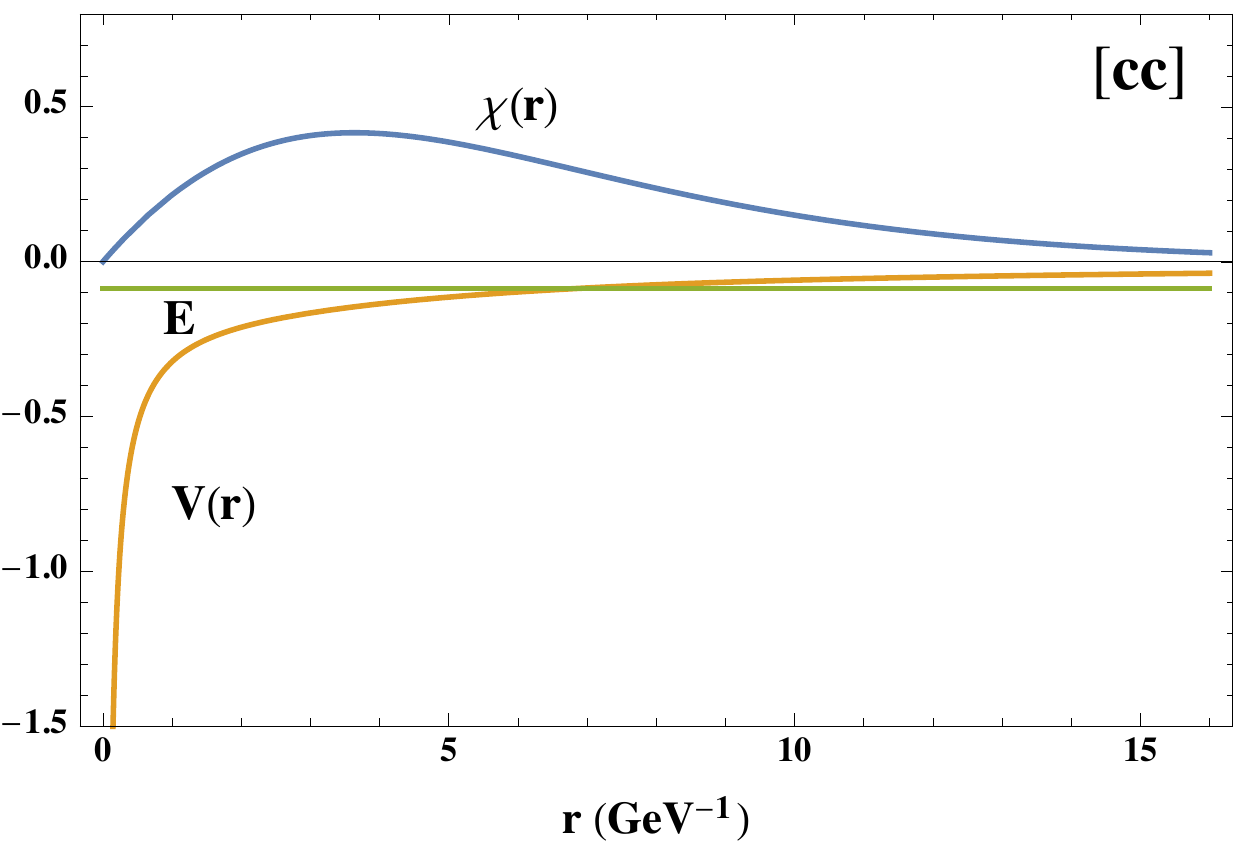}
 \end{minipage}%
\caption{\footnotesize Left Panel: BO potential, eigenfunction and eigenvalue $(bb)_{\bar{\bm 3}}\bar q\bar q$ tetraquark. Right Panel: same for $(cc)_{\bar{\bm 3}}\bar q\bar q$.  \label{qq3}}
\end{figure}
%@@@@@@@@@@@@@@@@@@@@@@@@@@@@@@@@@@@@@@@@@@

The BO potential, wave function and eigenvalue for the $bb$ pair in color  ${\bar{\bm 3}}$ and the one-gluon exchange couplings are reported in Fig.~\ref{qq3}. There is a bound tetraquark with a tight $bb$ diquark, of the kind expected in the constituent quark model~\cite{Karliner:2017qjm,Eichten:2017ffp,Luo:2017eub}.

The BO potential in the origin is Coulomb-like and it tends to zero, for large $r_{AB}$, due to~\eqref{noconf}. 
The (negative) eigenvalue $E$ of the Schr\"odinger equation is the binding energy associated with the BO potential. The masses of the lowest tetraquark with $(bb)_{S=1},~(\bar q\bar q)_{S=0}$ and of the $B$ mesons are
\bea
&& M(T)=2(M_b + M_q) + E+\frac{1}{2}\kappa_{bb}-\frac{3}{2}\kappa_{qq}\\
&&M(B)=M_b + M_q -\frac{3}{2}\kappa_{b\bar q}
\eea
%with: $\kappa_{bb}=15$~MeV, $\kappa_{qq}= 98$~MeV, $\kappa_{b\bar q}= 23$~MeV~\cite{Maiani:2004vq} 
The hyperfine interactions are taken from Tab.~\ref{spin} and $E=-67$~MeV is the eigenvalue shown in Fig~\ref{qq3}(a) with  $\alpha_s(2M_b)=0.20$. 

%@@@@@@@@@@@@@@@@@@@@@@@@@@@@@@@@@@@@@@@@@@@@@@
\begin{table}[ht!]
\begin{center}
\begin{tabular}{|c|c|c|c|c|c|}
\hline
{\footnotesize $QQ^\prime\bar u\bar d$}  & {\footnotesize This work } & \cite{Karliner:2017qjm} & \cite{Eichten:2017ffp}& \cite{Luo:2017eub} & {\footnotesize Lattice QCD}  \\
\hline
{\footnotesize $cc\bar u\bar d $} &  $+7 (-10)$ &  $+140$  & $+102$    & $+39$  & $-23\pm 11$~\cite{Junnarkar:2018twb}\\
{\footnotesize $cb\bar u\bar d$} &  $-60 (-74) $ &  $\sim 0$  &  $+83$  & $-108$ &  $+8\pm 23$~\cite{Francis:2018jyb}  \\ \hline %% & & & &\\
{\footnotesize $bb\bar u\bar d$}  &  $-138 (-156)$ & $-170$  & $-121$  & $-75$ & 
$\begin{array}{cr}-143\pm 34&\text{\cite{Junnarkar:2018twb}}\\ -143(1)(3) & \text{\cite{Francis:2016hui}}\\
-82\pm 24 \pm 10& \text{\cite{Leskovec:2019ioa}}\\
 \end{array}$\\
%\begin{pmatrix} $-143\pm 34$~\cite{Junnarkar:2018twb}\\ $-143(1)(3)$~\cite{Francis:2016hui}\end{pmatrix} \\
\hline
\end{tabular}
\end{center}
\caption{\footnotesize $Q$ values in MeV for decays into meson+meson+$\gamma$ obtained with string tension $1/4 \,k$ in Eq.~\eqref{orbitpot}, in parentheses with string tension  $k$. Models in~\cite{Karliner:2017qjm,Eichten:2017ffp,Luo:2017eub} are different elaborations of the constituent quark model we use throughout this paper, more details are found in the original references. In the last column the lattice QCD results~\cite{Junnarkar:2018twb,Francis:2018jyb,Francis:2016hui,Leskovec:2019ioa}.
 }\label{tab}
\end{table}
The $Q$-value for the decay $T\to 2 B +\gamma$ is then
\be
%&&
Q_{bb}=E+\frac{1}{2}\kappa_{bb}-\frac{3}{2}\kappa_{qq}+3~\kappa_{b\bar q}=-138 (-156)~{\rm MeV}
\label{eqe}
\ee
for the string tension \eqref{kbqbar3} (in parenthesis with string tension $k$).

Results for $Q_{cc,bc}$ are reported in Tab.~\ref{tab} using the values of $\alpha_S$ in  \eqref{bb&bc}.

Eq.~\eqref{eqe} underscores the result obtained by Eichten and Quigg~\cite{Eichten:2017ffp} that the $Q$-value goes to a negative constant limit for $M_Q\to \infty$: $Q= -150$~MeV$+{\cal O}(1/M_Q)$. 

{\bf \emph{Double beauty tetraquarks: $\bm b\bm b$ in $\bm 6$}}.
Color charges are given in \eqref{bqbar6} and 
\be 
k(b\bar q)=\frac{5}{8}~ k \label{kbqbar6}
\ee
The situation is entirely analogous to the $H_2$ molecule, with two identical, repelling light particles. 
%@@@@@@@@@@@@@@@@@@@@@@@@@@@@@@@@@@@@@@@@@
\begin{figure}[htb!]
   \centering
 \begin{center}
   \includegraphics[width=4.2truecm]{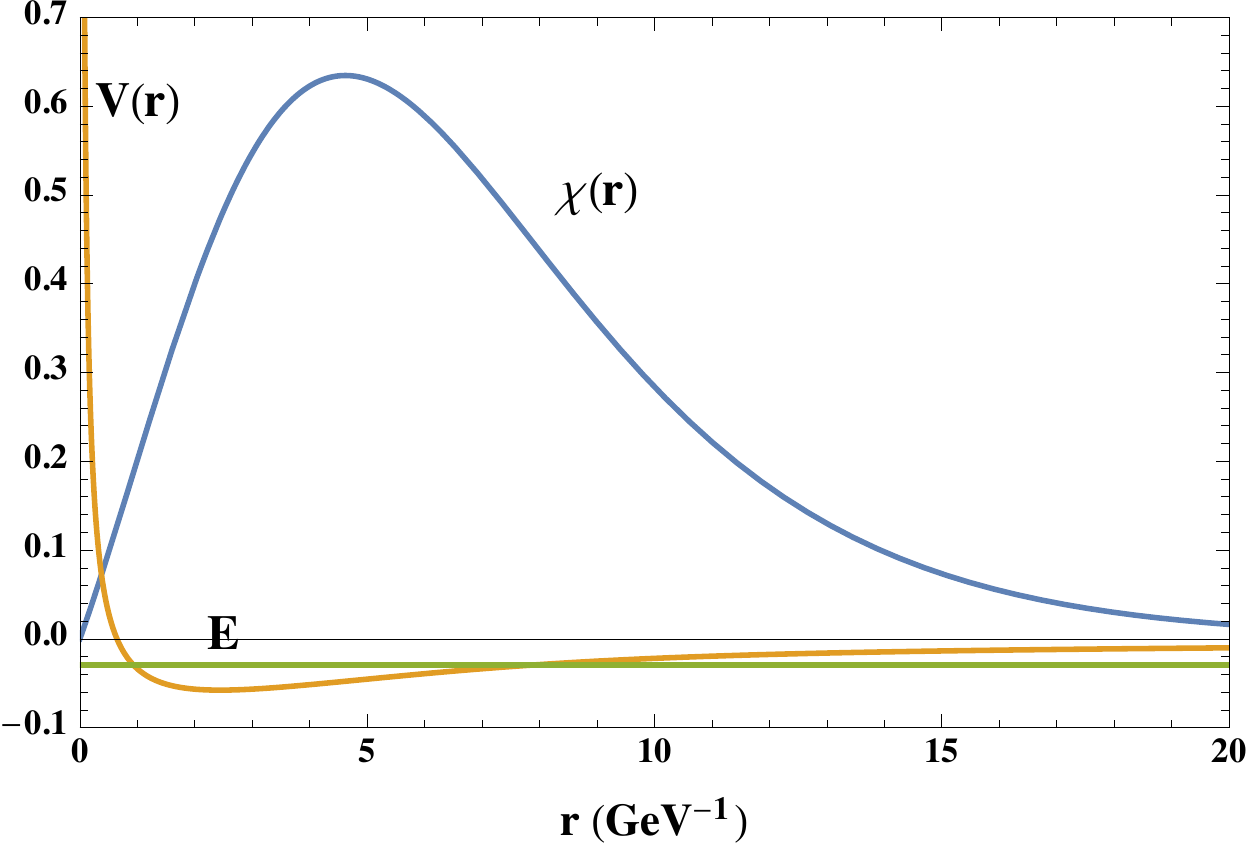}
 \end{center}
\caption{\footnotesize   A shallow bound state might be present in the color $\bm 6$ channel.  \label{nqq6}}
\end{figure}
%@@@@@@@@@@@@@@@@@@@@@@@@@@@@@@@@@@@@@@@@@@
For the orbital $b\bar q$, we find $A=0.43~{\rm GeV}$ and $\langle H\rangle_{\rm min}=0.72$~GeV.
The BO potential  with the one-gluon exchange parameters admits a very  shallow  bound state with $E=-30$~MeV, quantum numbers:  $(bb)_{{\bm 6},S=0}$ and $(\bar q\bar q)_{{\bar {\bm 6}},S=0, I=1}$, $J^{PC}=0^{++}$, and charges~$-2,-1,0$.

As shown in Fig.~\ref{nqq6}, the potential is so shallow as to raise doubts whether a bound tetraquark will indeed result. We register nonetheless the $Q$-value for the decay $T\to B\bar B$. For the string tension \eqref{kbqbar6} we find:
\be
Q_{bb}=E-\frac{3}{2}\kappa_{bb}-\frac{3}{2}\kappa_{qq}+3~\kappa_{b\bar q}=-131(-133)~{\rm MeV}
\ee
in parenthesis the result with string tension $k$.

\section{Summary of Results}\label{summ}

The present paper gives an extensive discussion of doubly heavy hadrons, baryons and tetraquarks, within the Born-Oppenheimer (BO) approximation.  The paper is an expansion of the shorter communication~\cite{noiprd}, with   the discussion of doubly heavy baryons added, a case where we can compare directly theory to experimental results~\cite{Aaij:2018gfl}. 
%We hope to complete in a future publication the next logical step, the case of doubly heavy pentaquarks. 

In analogy with the QED treatment~\cite{pauling} of the $H_2^+$ ion (the analog of a doubly heavy baryon)  and the $H_2$ molecule (analog of a doubly heavy tetraquark), we start our discussion from orbitals: two body, heavy-light,  quark-quark  or quark-antiquark lumps held together by the QCD Coulomb-like interaction plus a linear confining term with the appropriate string tension. 

The wave functions of the orbitals, obtained from the two body Schr\"odinger equation, are taken as zeroth order approximation of the light constituents wave function inside the hadron. QCD Coulomb-like interactions with the other constituents of the light quarks or antiquarks inside the orbitals are treated as perturbations, to obtain the first order BO potential that goes into the Schr\"odinger equation of the heavy constituents.

The non-abelian nature of QCD produces a number of peculiarities. Given that the hadron is a color singlet and given the representation of the heavy constituents, one can deduce, for each pair, the coefficient of the Coulomb-like interaction and the strenght of the string tension. The pair forming an orbital, except for the case of the baryon, is general  in a superposition of color representations with the same triality, {\it e.g.} ${\bar {\bm 3}}$ and ${\bm 6}$. Orbitals with non-vanishing triality have to be confined and we add to the BO potential the appropriate linearly rising potential. Triality zero orbitals are not confined, as discussed in Sect.~\ref{q_int} and ~\cite{Bali:2000gf}, and the BO potential vanishes for large separation of the heavy constituents.

A feature of the QCD Cornell potential, Sect.~\ref{q_int}, is that it contains an additive constant $V_0$ that in charmonium physics is determined from one physical mass of the spectrum.  We are able to determine $V_0$ (i) in the baryon case from a boundary condition related to the heavy quark-diquark symmetry~\cite{Savage:1990di, Brambilla:2005yk,Fleming:2005pd}, Sect.~\ref{doubch}, and (ii) in $QQ\bar q\bar q$ tetraquarks from the condition that, at infinity, the potential gives rise to a meson-meson$^*$ pair, Sect.~\ref{bbtetra}. For this reasons, we get in these two cases, an absolute prediction of their mass, which can be compared with the experimental value in the case of~$\Xi_{cc}$,  and which allows us to judge about the stability of $bb\bar q\bar q$ against strong or electromagnetic (e.m.) decays into $D B^*$ or $D B+ \gamma$.

On the other hand, $V_0$ remains undetermined for $Q\bar Q q\bar q$ tetraquarks and orbitals with non vanishing triality and the hadron mass cannot be predicted, at least for the ground state. However, the $Q\bar Q$ wave function provides interesting information on the tetraquark internal structure, with significant  phenomenological implications.

We now summarize the results of Sects.~\ref{doubch} to \ref{bbtetra}

{\bf \emph{Doubly heavy baryon.}} 
Our results are summarized in Tab.~\ref{dhbar}, fourth column. We find $M(\Xi_{cc})_{{\rm Th}}=3652^{+17}_{-7}\label{xicc}~{\rm MeV}$
to be compared with the LHCb value~\cite{Aaij:2018gfl} $M(\Xi_{cc})_{{\rm Expt}}=3621.2\pm 0.7~{\rm MeV}$. The difference is within the theoretical uncertainty of our approach, see Eq.~\eqref{err1}.
For the heavier baryons, our results differ from the results in Ref.~\cite{Karliner:2014gca,Karliner:2018hos} by 50 and 150 MeV for $bc$ and $bb$ baryons, respectively. Recent lattice QCD results~\cite{Mathur:2018epb,Mathur:2018rwu}, where available, are intermediate between us and~\cite{Karliner:2014gca,Karliner:2018hos}, see Tab.~\ref{dhbar}.

Overall, the general consistency of results derived by alternative routes with themselves and with the experimental value is very encouraging. Experimental results on heavier baryons will allow a more significant comparison and are eagerly awaited.

{\bf \emph{Hidden charm tetraquark: ${\bm c}{\bm q}$ orbitals.}} 
The interaction between the light quarks, $q$ and $\bar q$ is repulsive. Combined with the existence of a raising confining potential between the orbitals, this leads to envisage two regimes, exemplified in Figs.~\ref{uno}(a), (b). 

For the low value of the repulsive coupling, $+1/6 \,\alpha_S\sim 0.05$, implied by one gluon exchange, the equilibrium configuration obtains for $c$ and $\bar c$ relatively close to each other, in a quarkonium adjoint meson configuration~\cite{braatenBO,Brambilla:2017uyf}, see Fig.~\ref{uno}(a).

Increasing the repulsion, orbitals are split apart and equilibrium obtains for a diquark-antidiquark configuration,~\ref{uno}(b), with well separated diquarks. 
As an example, letting $+1/6 \,\alpha_S\sim 0.05\to 3.3$ in Eq.~\eqref{lamccbar}, diquarks are separated by a potential barrier and there are two different lenghts: the diquark radius $R_{qc}\sim 0.4 $~fm and the total radius $R\sim 1.5$~fm.
A dominant, non-perturbative $q \bar q$ repulsion plus confinement gives the dynamical basis to the emergence of the repulsive barrier between diquarks and antidiquarks suggested in~\cite{Maiani:2017kyi}. The need to tunnel under the barrier explains why decays into charmonia occur at a lower rate with respect to decays into  open charm mesons, as observed in $X$ and $Z$ resonances.  Diquark-antidiquark separation may also be the reason why charged partners of the $X$ have not (yet) been observed and there is an almost degenerate doublet of $X^{0}_{u,d}$ neutral states~\cite{Maiani:2017kyi,Esposito:2018cwh}.

{\bf \emph{Hidden charm tetraquark: ${\bar {\bm c}}{\bm q}$ orbitals.}}
The BO potential goes to $+\infty$ at zero separation, due to $c\bar c$ repulsion, and it vanishes at infinity, due to the zero triality of orbitals. The existence of a minimum is not guaranteed. The situation is shown in Figs.~\ref{due}(a),(b). For the one gluon exchange parameters, there is indeed one minimum, Fig.~\ref{due}(a), and a second tetraquark, in the quarkonium adjoint meson configuration. 

If  the $q\bar q$ repulsion is increased, letting {\it e.g.} $+1/6 \,\alpha_S\sim 0.05$ to a value $>1$, there is no mimimum, Fig.~\ref{due}(b). The lack of a second resonance with the same features of, but well separated from $X(3872)$, would speak in favour of Figs.~\ref{uno}(b) and \ref{due}(b), supporting the enhancement of $q\bar q$ repulsion.
%$$$$$$$$$$$$$$$$$$$$$$$$$$$$$$$$$$$$$$$$$$$$$$$$$$$$$$$

{\bf \emph{Double heavy tetraquarks: $(QQ)_{{\bar {\bm 3}}}$.}}
Our results for the $Q$-value of the lowest $[bb]$ tetraquark against decays into $DB^*+\gamma$ are shown in Tab.~\ref{tab} and found to compare well with previous estimates done with quark model, Ref.~\cite{Karliner:2017qjm,Eichten:2017ffp,Luo:2017eub}  and, remarkably, with Lattice QCD results~\cite{Junnarkar:2018twb,Francis:2018jyb,Francis:2016hui,Leskovec:2019ioa}, where available. 

Given the error estimate following Eq.~\eqref{err2}, we support the proposal  that the lowest $[bb]$ and perhaps $[bc]$ tetraquarks  may be stable against strong and electromagnetic decays~\cite{Karliner:2017qjm,Eichten:2017ffp}, see also~\cite{Ali:2018xfq,Ali:2018ifm}.

{\bf \emph{Double heavy tetraquarks: $(QQ)_{ {\bm 6}}$.}} The $V_{BO}$ pptential for ${bb}$ has a repulsvi behaviour t the origin and it vanishes at large separations. with a very shallow minimum. 

The binding energy $E=-30$~MeV is at the limit of our visibility. If it exists, the bound state would make a second $bb$ tetraquark, possibly stable. Its existence needs confirmation by lattice QCD calculations.

\section{Conclusions}\label{concl}
The BO approximation gives a new insight on multiquark hadron structure and provides new opportunities for theoretical progress in the field of exotic resonances.

The restriction to a perturbative treatment followed here is, at the moment, a necessity for any analytical approach. Nonetheless, the consistency of the results we have found for doubly heavy baryons and doubly heavy tetraquarks with lattice QCD calculations seems to show that the perturbative approach is sufficiently robust (as it was for the Hydrogen ion and molecule) to provide useful, quantitative indications. 

A critical case, where non perturbative calculations are called for is in the $Q\bar Q q\bar q$ tetraquarks. As we have shown here, the strength of $q\bar q$ repulsion is the critical parameter to determine the internal configuration of the tetraquark, from a quarkonium adjoint meson to a diquark-antidiquark configuration. The latter configuration is indicated by the pattern of decay modes of $X(3872)$ and is compatible with the existence of charged partners of the $X(3872)$ not to be observed in open charm decays but only in final states containing charmonia, $X^\pm\to \rho^\pm\,J/\psi$. The $B$ meson may have smaller branching fraction than expected for decays that involve the charged $X$ and this requires some dedicated experimental effort to go beyond the bounds which have been set years ago.

 Non-perturbative investigations along these lines should be provided by lattice QCD, following the growing interest shown for doubly heavy tetraquarks.

\acknowledgements
We are grateful for hospitality by the T. D. Lee Institute and Shanghai Jiao Tong University where this work was initiated. 
We acknowledge interesting discussions with A. Ali, A. Esposito, A. Francis,  M. Karliner, R. Lebed, N. Mathur, A. Pilloni and W. Wang.

\appendix
%@@@@@@@@@
\section{ QED orbitals and molecules}\label{qedmol}

We  review here the Born-Oppenheimer  approximation for the hydrogen molecule and sketch the perturbative method starting from the hydrogen orbitals~\cite{pauling} which provides the basis of our treatment of heavy-light tetraquarks in QCD.

The Hamiltonian of two protons in ${\bm x}_A$ and ${\bm x}_B$ and two electrons in ${\bm x}_1$ and ${\bm x}_2$ is
\bea 
&&H=\sum_{A,B}\frac{P_i^2}{2M}+\sum_{1,2}\frac{p_i^2}{2m}+\alpha\left(\frac{1}{|{\bm x}_A-{\bm x}_B|}\right)-\nonumber \\
&&-\alpha\left(\frac{1}{|{\bm x}_A-{\bm x}_1|}+\frac{1}{|{\bm x}_B-{\bm x}_2|}+(1\leftrightarrow 2)\right)+\nonumber \\
&&+\alpha\frac{1}{|{\bm x}_1-{\bm x}_2|}= H_{AB}+H_{A,1}+ H_{B,2} + H_{\rm pert}
\eea
where
\bea
&&H_{AB}=\sum_{A,B}\frac{P_i^2}{2M}+\alpha\left(\frac{1}{|{\bm x}_A-{\bm x}_B|}\right)\nonumber \\
&&H_{A,1}=\frac{p_1^2}{2m}-\alpha\frac{1}{|{\bm x}_A-{\bm x}_1|}\nonumber \\
&& H_{B,2}={\rm same~with:}~A\to B,~1\to 2\nonumber \\
&&H_{\rm pert}=-\alpha\left(\frac{1}{|{\bm x}_A-{\bm x}_2|}+\frac{1}{|{\bm x}_B-{\bm x}_1|}\right)+\nonumber \\
&&+\alpha\frac{1}{|{\bm x}_1-{\bm x}_2|}
\eea
We denote  by $\psi(x)$ the lowest energy eigenfunction of $H_{A,1}$ and by $\phi(x)$ the similar eigenfunction of $H_{B,2}$, both being real functions. Since they belong to two different Hamiltonian, $\psi(x)$ and $\phi(x)$ are not orthogonal and we denote by $S$ the overlap function
\be
S(r_{AB})=\int~d^3x~ \psi(x)\phi(x)
\ee
with $r_{AB}=|{\bm x}_A-{\bm x}_B|$. $\psi$ and $\phi$ are usually called the {\it orbitals} of the  $H_2$ molecule. Neglecting $H_{\rm pert}$, there are two degenerate lowest energy eigenstates, namely $\psi(x_1)\phi(x_2)$ and $\psi(x_2)\phi(x_1)$, which may be combined in the symmetric or antisymmetric combinations. When $H_{\rm pert}$ is turned on, the antisymmetric combination turns out to have a higher energy and we restrict to the symmetric combination ($\psi$ and $\phi$ normalised to unity):
\be
f_0=\frac{ \psi(x_1)\phi(x_2)+\psi(x_2)\phi(x_1)}{\sqrt{2(1+S^2)}}\label{f0}
\ee
with energy
\be
E_0=2E_H=-\alpha^2 m
\ee
i.e. twice the Hydrogen ground level. Electrons being fermions, the symmetric combination~\eqref{f0} is associated with electron spins in the singlet combination, $S=0$.
 
To first order in $H_{\rm pert}$ we find~\cite{pauling}:
\bea
&&E=E_0 + \Delta E(r_{AB})\nonumber\\
&&\Delta E=\langle f_0|H_{\rm pert}|f_0\rangle=\nonumber \\
&&=\frac{\alpha}{(1+S^2)}\left[ -2(I_1+S I_2)+I_4+I_6 \right]
\eea
$I_1$ to $I_6$ as functions of $r_{AB}$ are defined as:
\bea
&& I_1=\int~d^3x~\psi(x)^2\frac{1}{|{\bm x}_B-{\bm x}|};\nonumber \\
&& I_2=\int~d^3x~\psi(x)\phi(x)\frac{1}{{|\bm x}_A-{\bm x}|};\nonumber \\
&& I_4=\int~d^3x d^3x~\psi(x)^2~\phi(y)^2~\frac{1}{r};\nonumber \\
&& I_6=\int~d^3x d^3x~[\psi(x)\phi(x)][\psi(y)\phi(y)]~\frac{1}{r}
\eea
with $r=|{\bm x}-{\bm y}|$. Explicit expressions of the integrals are given in~\cite{pauling}.

The Born-Oppenheimer potential is
\be
V_{BO}(r_{AB})= +\alpha~\frac{1}{r_{AB}}- \alpha^2 m +\Delta E(r_{AB})
\ee
The potential diverges to $+\infty$ for $r_{AB}\to 0^+$ and tends to $-\alpha^2 m$ (the energy of two hydrogen atoms), for $r_{AB}\to \infty$. A numerical evaluation of the previous formulas shows that the potential has one minimum for:
\bea
&&r_{\rm min}\sim1.5~ (\alpha m)^{-1}= 0.79 ~{\rm \AA}~(0.76 ~{\rm \AA})~\nonumber \\
&& [V_{BO}]_{\rm min}\sim  0.23 ~E_H= 3.1~ {\rm eV}~(4.4~ {\rm eV})\nonumber
\eea
which compare favourably with the experimental numbers given in parentheses.

Computed along the same lines, the BO potential for the antisymmetric combination (and electrons in the triplet state) shows no minimum.

\section{Fierz identities}\label{fierz}

The basic Fierz identity, in $SU(3)_c$, reads:
\be
\delta^\gamma_\alpha \delta^\delta_\beta=\frac{1}{3}\delta^\gamma_\beta \delta^\delta_\alpha +\frac{1}{2}(\lambda^A)^\gamma_\beta (\lambda^A)^\delta_\alpha
\ee
where from we derive
\bea
&&\delta^\gamma_\alpha \delta^\delta_\beta-\delta^\gamma_\beta \delta^\delta_\alpha=-\frac{2}{3}\delta^\gamma_\beta \delta^\delta_\alpha +\frac{1}{2}(\lambda^A)^\gamma_\beta (\lambda^A)^\delta_\alpha \label{threebar}\\
&&\delta^\gamma_\alpha \delta^\delta_\beta+\delta^\gamma_\beta \delta^\delta_\alpha=+\frac{4}{3}\delta^\gamma_\beta \delta^\delta_\alpha +\frac{1}{2}(\lambda^A)^\gamma_\beta (\lambda^A)^\delta_\alpha \label{six}
\eea

Saturating with the products $q^\alpha Q^\beta \bar Q_\gamma \bar q_\delta$, we obtain the identities:
\bea
&&(\bar Q q)(\bar q Q)-(\bar Q Q)(\bar q q)=\nonumber \\
&&=-2 \frac{(\bar Q Q)(\bar q q)}{3}+ 2 \sqrt{2}\frac{(\bar Q \lambda^A Q)(\bar q \lambda^A q)}{4 \sqrt{2}}\nonumber \\
&&(\bar Q q)(\bar q Q)+(\bar Q Q)(\bar q q)=\nonumber \\
&&=4 \frac{(\bar Q Q)(\bar q q)}{3}+ 2 \sqrt{2}\frac{(\bar Q \lambda^A Q)(\bar q \lambda^A q)}{4 \sqrt{2}}\nonumber
\eea
factors in the denominators are introduced to have quadrilinear forms normalised to unity 
\footnote{for an expression of the form $T\otimes T^\prime$ with $T$ and $T^\prime$ matrices in color space, we require ${\rm Tr}(T T^\dagger)={\rm Tr}(T^\prime T^{\prime \dagger})=1$. If we have a sum $\sum_AT^A\otimes T^{\prime A},~A=1,\dots N$, with each term normalised to unity, we divide by an additional factor $\sqrt{N}$.}
.

In terms of normalised kets, we have
\bea
&&|(Qq)_{\bar{\bm 3}} (\bar Q \bar q)_{\bm 3}\rangle_1=\nonumber \\
&&=\frac{1}{\sqrt{3}}|(\bar Q Q)_{\bm 1} (\bar q q)_{\bm 1}\rangle_{\bm 1}-\sqrt{\frac{2}{3}}|(\bar Q Q)_{\bm 8} (\bar q q)_{\bm 8}\rangle_{\bm 1}\nonumber\\
&&|(Qq)_{\bm 6} (\bar Q \bar q)_{{\bar {\bm 6}}}\rangle_1=\nonumber \\
&&=\sqrt{\frac{2}{3}}|(\bar Q Q)_{\bm 1} (\bar q q)_{\bm 1}\rangle_{\bm 1}+\frac{1}{\sqrt{3}}|(\bar Q Q)_{\bm 8} (\bar q q)_{\bm 8}\rangle_{\bm 1}\nonumber
\eea
The combination with $Q \bar Q$ in pure octet is therefore
\bea
&&T=|(\bar Q Q)_{\bm 8} (\bar q q)_{\bm 8}\rangle_{\bm 1}=\nonumber \\
&&=\sqrt{\frac{2}{3}} |(Qq)_{\bar{\bm 3}} (\bar Q \bar q)_{\bm 3}\rangle_1-\frac{1}{\sqrt{3}}|(Qq)_{\bm 6} (\bar Q \bar q)_{{\bar {\bm 6}}}\rangle_1\nonumber
\eea
so that
\be
\lambda_{Qq}=\lambda_{\bar  Q\bar q}=\left[\frac{2}{3}\left(-\frac{2}{3}\right)+\frac{1}{3}\frac{1}{3}\right]\alpha_S=-\frac{1}{3}\alpha_S
\ee

Saturating (\ref{threebar}) and (\ref{six}) with the combination: $Q^\alpha q^\beta \bar Q_\gamma \bar q_\delta$, we express the diquark-antidiquark states in terms of the bilinears with the pairs $\bar Q q$ and $\bar q Q$ and finally express the latter in terms of the state $T$:
\be
T=\sqrt{\frac{8}{9}}|(\bar Q q)_{\bm 1}(\bar q Q)_{\bm 1}\rangle_{\bm 1}-\frac{1}{\sqrt{9}}|(\bar Q q)_{\bm 8}(\bar q Q)_{\bm 8}\rangle_{\bm 1} \label{Qqbar}
\ee
and 
\be
\lambda_{\bar Q q}=\lambda_{\bar q Q}=-\frac{7}{6}\nonumber
\ee

\section{Mass and mixing of $\Xi_{cb}$ and $\Xi_{cb}^\prime$}\label{xispin}

For identical $cc$ or $bb$ flavors, color antisymmetry and Fermi statistics require the pair to be in spin $1$ and there is only one state for total spin $J=1/2$. In the case of $cb$, there are two states with $J=1/2$ and $S_{cb}=0,1$. It is customary to classify the states according to the spin of the lighter quarks, namely
\be
[\Xi_{cb}]_0=|(qc)_0;b\rangle_{1/2}\quad\quad[\Xi^\prime_{cb}]_0=|(qc)_1;b\rangle_{1/2}\label{barestat}
\ee
where the subscript $0$ on brackets refer to states before mixing and the subscript $0,1$ inside kets refer to the total spin  of the lighter pair.

The hyperfine Hamiltonian is
\be
H_{\rm hf}=2\kappa_{qc}({\bm s}_q \cdot {\bm s}_c)+2\kappa_{qb}({\bm s}_q \cdot {\bm s}_b) +2\kappa_{cb}{(\bm s}_c \cdot {\bm s}_b)
\ee
and to compute the matrix elements we need to know what is the spin if the $qb$ and $cb$ pairs in the states~\eqref{barestat}, see {\it e.g.}~\cite{book}.

An elementary calculation gives (we drop for simplicity the subscript $cb$):
\bea
&&\Xi_0=\frac{\sqrt{3}}{2}|[(qb)_1 c]_{1/2}\rangle+\frac{1}{2}|(qb)_0 c\rangle=\nonumber \\
&&=-\frac{\sqrt{3}}{2}|[(cb)_1 u]_{1/2}\rangle-\frac{1}{2}|(cb)_0 u\rangle\nonumber \\
&&\Xi^\prime_0=-\frac{1}{2}|[(qb)_1 c]_{1/2}\rangle+\frac{\sqrt{3}}{2}|(qb)_0 c\rangle=\nonumber \\
&&=+\frac{1}{2}|[(cb)_1 q]_{1/2}\rangle-\frac{\sqrt{3}}{2}|(cb)_0 q\rangle
\eea
Scalar products $({\bm s}_i \cdot {\bm s}_j)$ commute with the total spin ${\bm S}_{ij}$ and we find
\be
\langle\Xi_0|{\bf s}_q\cdot {\bf s}_c|\Xi_0\rangle=-\frac{3}{2}\quad\quad \langle\Xi^\prime_0|{\bf s}_q\cdot {\bf s}_c|\Xi^\prime_0\rangle=+\frac{1}{2}\notag
\ee
and
\bea
&& \langle\Xi_0|{\bf s}_q\cdot {\bf s}_b|\Xi_0\rangle=\langle\Xi_0|{\bf s}_c\cdot {\bf s}_b|\Xi_0\rangle=0\notag\\
&& \langle\Xi_0^\prime|{\bf s}_q\cdot {\bf s}_b|\Xi_0^\prime\rangle=\langle\Xi_c^\prime|{\bf s}_c\cdot {\bf s}_b|\Xi_c^\prime\rangle=-1\notag\\
&&\langle\Xi_0^\prime|H_{\rm hf}|\Xi_0\rangle=\frac{\sqrt{3}}{2}(\kappa_{qb}-\kappa_{cb})\notag
\eea
 The mixing matrix, in the ($\Xi_0, \Xi_0^\prime$) basis is
 \be
 M(\Xi)=\left(\begin{array}{cc} -\frac{3}{2}\kappa_{qc}  &   \frac{\sqrt{3}}{2}(\kappa_{qb}-\kappa_{cb}) \\
  \frac{\sqrt{3}}{2}(\kappa_{qb}-\kappa_{cb})& +\frac{1}{2}\kappa_{qc}-\kappa_{qb}-\kappa_{cb} \end{array}\right)
 \ee
Numerically, we use Tabs.~\ref{mas} and \ref{spin}. Noting that $\kappa_{ij} \propto (M_i M_j)^{-1}$, see~\cite{book}, we take
 \be
 \kappa_{bc}=\sqrt{\kappa_{cc}\kappa_{bb}} \notag
 \ee
to obtain the eigenvalues: ($-35,-2.9$) MeV and the $\Xi_{cb}$ and $\Xi^\prime_{cb}$ masses reported in Tab.~\ref{dhbar}.
 
\newpage
%@@@@@@@@@@@@@@@@@@@@@@@@@@@@@

\end{document}